\documentclass[preprint,aps,tightenlines]{revtex4}

\usepackage{epsf}
\usepackage{float}
\usepackage{epsfig}

\begin{document}

\title{Analysis of a Parametrically Driven Pendulum}

\author{Randy Kobes}
\email{randy@theory.uwinnipeg.ca}
\altaffiliation{ Winnipeg Institute for
        Theoretical Physics, Winnipeg, Manitoba, R3B 2E9 Canada}
\affiliation{Department of Physics, University of Winnipeg, Winnipeg, 
        Manitoba, R3B 2E9 Canada}

\author{Junxian Liu}
\email{jliu@tikva.chem.utoronto.ca}
\altaffiliation{Department of Physics, University of Winnipeg, Winnipeg, 
        Manitoba, R3B 2E9 Canada}
\affiliation{Department of Chemistry, University of Toronto,
        Toronto, Ontario M5S 3H6 Canada}

\author{Slaven Pele\v{s}}
\email{peles@theory.uwinnipeg.ca}
\affiliation{Department of Physics, University of Manitoba, Winnipeg, 
        Manitoba, R3T 2N2 Canada}

\begin{abstract}
We study in this paper the behavior of 
a periodically driven nonlinear mechanical system. 
Bifurcation diagrams are found which locate regions of quasiperiodic, 
periodic and chaotic behavior within the parameter space of the
system. We also conduct a symbolic analysis of the model,
which demonstrates 
that the symbolic dynamics of two-dimensional maps can be applied
effectively to the study of ordinary differential equations 
in order to gain global knowledge about them.
\end{abstract}

\maketitle

\section{Introduction}
In this paper we conduct an investigation of the behavior of 
a periodically driven nonlinear mechanical system. 
The system is of interest both from the point of view
of the relatively rich structure it possesses and also
because it is fairly easy to construct a mechanical
model of it. 
\par
We begin with some standard techniques used to analyze
such systems. 
Bifurcation diagrams are plotted which locate regions of quasiperiodic, 
periodic and chaotic behavior within the parameter space of the
system. Dimensions of the attractor are 
calculated for different values of control parameters and compared 
with dimensions estimated from Lyapunov exponents. Numerical results obtained 
are in agreement with the Kaplan-Yorke conjecture. Multiple attractors are 
found in different regions in parameter space. The most interesting is 
the coexistence of a chaotic attractor and a limit cycle, which implies that 
initial conditions may determine whether the system will exhibit periodic or 
chaotic behavior. Boundaries of the basins of attraction appear to be smooth 
for high dissipation in the system, and they show fractal structure as 
the dissipation decreases. Creation and destruction of multiple attractors 
in this model is caused by tangent bifurcations and crisis phenomena. This 
illustrates how even a simple nonlinear mechanical system may exhibit 
fairly complex behavior and posses a variety of chaotic features.
\par
We also conduct a symbolic analysis of the system.
Symbolic dynamics provides almost the only rigorous approach
to study the motion of dynamic systems. Such an analysis of 
one-dimensional maps on the interval is well understood \cite{mss,h89}.
For the simplest case of the unimodal map, a binary generating partition
may be introduced by splitting the interval at the critical point. 
By assigning the  letter
$R$ or $L$ to a point of an orbit, depending whether it falls to the right
or left side 
of the critical point, the orbit can be encoded with a symbolic sequence.
The kneading sequence, which is the forward sequence 
of the critical value, determines all the admissible sequences of the map.
By means of the kneading theory, all sequences are well ordered. The kneading 
sequence is the greatest. There is no allowed sequence which has a  
shifted sub-sequence greater than the kneading sequence.

The symbolic dynamics of the unimodal map can be extended to that of
two-dimensional (2D)
maps. Two much-studied 2D models are the H\'enon map 
\cite{henon} and its
piecewise 
linear version, the Lozi map \cite{lozi,zl94}. First of all, one needs to
construct a 
`good' binary partition for them. In Ref.\cite{gk85}, 
by considering all `primary'
homoclinic tangencies of the H\'enon map, a method for determining a
partition line was 
proposed. Once the binary partition is determined, any orbit may be
associated with a doubly infinite symbolic sequence 
$S = \cdots s_{\overline{m}} \cdots s_{\overline{2}}s_{\overline{1}}\bullet 
s_0 s_1 \cdots s_n \cdots$ where $s_0$ indicates the code of the initial
point. The 
forward sequence $\bullet s_0 s_1 \cdots s_n \cdots $ and backward sequence
$\cdots s_{\overline{m}} \cdots s_{\overline{2}}s_{\overline{1}}\bullet$
correspond to
the forward and backward orbits of the initial point, respectively. In
Ref.\cite{cgp} 
the ordering
rules for forward and backward sequences were discussed, and by introducing
a metric
representation of sequences the symbolic plane was constructed. It was
pointed out that
every primary homoclinic tangency cuts out a rectangle of forbidden
sequences in the
symbolic plane. The rectangles so deleted build up a pruning front which is
monotonic across
half the symbolic plane. By generalizing stable and unstable manifolds of
an unstable fixed
or periodic point to forward and backward foliations of any point \cite{gu87},
homoclinic
tangencies are generalized to tangencies between the two classes of 
foliations.
In terms of the generalization, we can make symbolic analysis of the orbits
not only in but
also out of attractor, including transient orbits.
The extension of symbolic dynamics of maps from 1D to 2D is made by
decomposing
a 2D map into two 1D maps based on forward and backward foliations.
The coupling between the two 1D maps is described by the pruning front or
the
symbolic representation of the partition line. An attractor is associated
with 
backward foliations. 

It has been convincingly demonstrated that a properly
constructed two-dimensional symbolic dynamics, being a coarse-grained 
description, provides a powerful tool to capture global, topological 
aspects of low-dimensional dissipative systems of ordinary 
differential equations (ODEs)
\cite{lz95,lzh96,xzh95,zl95,lwz96,f95,fh96,zl97,hlz98,hao98,z9194}. 
Therefore, 
to have a further global understanding of the bifurcation and chaos
``spectrum'' in the system we have studied \cite{ito81},
i.e. the systematics of stable and
unstable periodic orbits at varying and fixed parameters, the types 
of chaotic attractors, etc., we shall carry out a full analysis of symbolic
dynamics for this mechanical model.
Based on the primary 
tangencies between forward and backward foliations in the Poincar\'e 
section the appropriate partition of the phase space is made and three
letters are used to describe the dynamics. From the ordering rules and 
metric representation of forward and backward sequences symbolic planes 
are constructed and the admissibility conditions for allowed sequences 
derived. Unstable allowed periodic orbits embedded in a chaotic attractor
are predicted through symbolic analysis and verified
numerically as well. In the parameter space period windows are located and
the corresponding period words are determined according to the
partitioning of phase portraits. 
\par
The paper is organized as follows.
In section Section \ref{p1} a 
brief description of the mechanical system and a derivation of 
the equations of motion is given. In Section \ref{p2}
chaos is detected, 
and regions of periodic, quasiperiodic and chaotic behavior are located 
within the parameter space. Lyapunov exponents and associated dimensions are 
calculated for certain attractors, and the Kaplan-Yorke conjecture is 
established. In section \ref{sec:cox} 
crisis phenomena and coexisting 
attractors are studied. We next turn to the symbolic
dynamics of the module.
In Section \ref{sec2}, the global dynamical behavior of the system while
parameters
vary are shown in phase space, and the partition lines are determined from
tangencies 
between forward and backward foliations in the Poincar\'e section. In
Section \ref{sec3}
by introducing a metric representation of forward and backward sequences
according 
to their ordering rule the symbolic plane is constructed, and the
admissibility 
condition of sequences is discussed. We then analyze the admissibility
of periodic sequences based on a finite number of points on the partition
lines, 
and give the numerical results of allowed periodic orbits in 
Section \ref{sec4}. We
also locate
the period windows for some range of parameters, assign
words to the stable period orbits based on the partitioning of phase space,
and discuss
their ordering properties as parameters change in Section \ref{sec5}.
In Section \ref{prop} we discuss some aspects of this
model as used in a mechanical propulsion device. 
Finally, in Section \ref{con} we present some conclusions.
\section{The Equation of Motion}
\label{p1}
\begin{figure}[H]
\begin{center}
        \leavevmode
        \epsfysize=7 cm
        \epsfbox{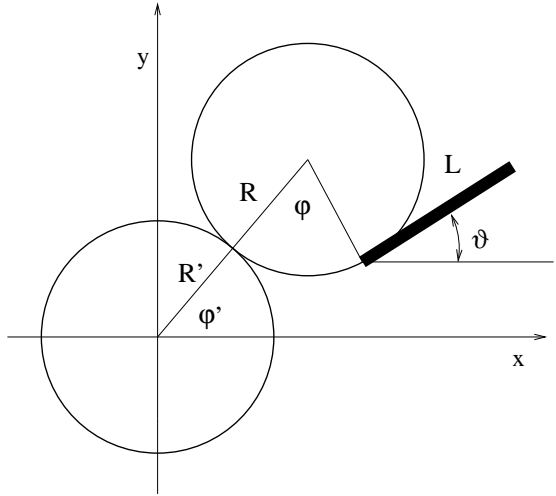}
        \label{fig:rod}
        \caption{Two gears and the rod}
\end{center}
\end{figure}

The system we study consists of two gears and a rod. The first one, 
the solar gear, has 
radius $R'$, and is fixed so neither its center of mass can move nor can
it rotate around its axis. The second, the planetary gear, is of 
radius $R$ and is attached to the first gear so when it rotates around 
its axis at the same time it goes around the fixed gear. 
We assume that the movable gear is powered by some device which keeps 
it moving with constant angular velocity $\omega_{d}$. When the planetary 
gear rotates for the angle $\phi$ around its axis it moves around solar 
gear for an angle $\phi'$ (Fig.~\ref{fig:rod}). 
The relation between two angles is
\begin{equation}
        R'\phi'=R\phi \label{eq:angles}
\end{equation}
A uniformly dense rod of length $L$ is at one end 
joined to the fringe of movable gear. The rod can rotate around the joint, 
assumed without friction. Both gears and the rod lie in a horizontal 
plane so the potential energy of the system is constant during the motion. 
We choose a coordinate system with origin at the center of the fixed gear. 
The system has two degrees of freedom, and a convenient choice of 
coordinates are $\phi'$, the angle that the second gear makes 
with $x$--axis, 
and $\vartheta$, the angle that the rod makes with the $x$--axis. 
Since the potential energy for the system is
constant as well as the 
angular velocity of the planetary gear, the Lagrangian for 
the system is determined by the kinetic energy of the rod only. A general 
expression for the kinetic energy for the rigid body that moves in two 
dimensions is
\begin{equation}
        T=\frac{1}{2} \int dm\,(\dot{x}+\dot{y}) \label{eq:kinetic}
\end{equation}
We may express the position coordinates for the mass element $dm$ over 
coordinates $\phi$ and $\vartheta$ as
\begin{equation}
\begin{array}{c}
        x=(R'+R)\cos \phi'-R\cos(\phi'+\phi)
                +l\cos \vartheta \\
        y=(R'+R)\sin \phi'-R\sin (\phi'+\phi)
                +l\sin \vartheta  \label{eq:coord}
\end{array}
\end{equation}
where $l$ is position of the mass element $dm$ at the rod. Deriving $x$ and 
$y$ over time and eliminating $\phi'$ rather than $\phi$, using 
(\ref{eq:angles}), we get:
\begin{equation}
\begin{array}{l}
        \dot{x}=-(1+r)R\dot{\phi}\sin r\phi+
                R(1+r)\dot{\phi}
                \sin\left[\left(1+r\right)\phi\right]-
                l \dot{\vartheta} \sin \vartheta \\
        \dot{y}=(1+r)R\dot{\phi}\cos r\phi-
                R\dot{\phi}\left(1+r\right)
                \cos\left[\left(1+r\right)\phi\right]+
                l \dot{\vartheta} \cos \vartheta  \label{eq:dxdy}
\end{array}
\end{equation}
where $r=R/R'$. Our initial assumption was that angular velocity 
of the planetary gear was constant $\dot{\phi}=\omega_{d}$. 
For simplicity, we will assume that initial condition is chosen so 
that $\phi=\omega_{d}t$. Substituting derivatives (\ref{eq:dxdy}) into 
(\ref{eq:kinetic}) and integrating over the whole rod, assuming its linear 
mass is constant, we get the kinetic energy, and therefore the
Lagrangian, to be:
\begin{equation}
        L = T = \frac{1}{2}I\dot{\vartheta}^{2}+\frac{1}{2}(1+r) 
        mLR\omega_{d}\dot{\vartheta} \left \{ \cos(\vartheta-r\phi)
        -\cos \left[\vartheta-\left(1+r\right)\phi\right]\right \}
        \label{eq:kinetic2}
\end{equation}
Here $I$ is the moment of inertia of the rod, and we dropped terms that can 
be written as a total time derivative of some 
function and therefore do not contribute to the Lagrangian. Substituting this 
Lagrangian into the general expression for the
Euler-Lagrange equations of motion of dissipative systems
\begin{equation}
        \frac{d}{dt}\frac{\partial L}{\partial \dot{q}_{\alpha}}-
                \frac{\partial L}{\partial q_{\alpha}}
                        =Q_{\alpha}, ~~~ \alpha = 1,...,s \label{eq:ELD}
\end{equation} 
and choosing $\vartheta$ and $\phi$ for coordinates we obtain the
equation of motion to be
\begin{equation}
        I\ddot{\vartheta}+\frac{1+r}{2}
        mLR\omega_{d}^{\,2} \left \{ r\sin(\vartheta-r\phi)-(1+r)
        \sin\left[\vartheta-(1+r)\phi \right] \right \} = Q_{\vartheta}
\end{equation}
where $Q_{\vartheta}$ is some friction force that acts upon the rod, 
and $I$ is its moment of inertia. If we assume that the friction, as
is commonly done for a pendulum, 
has the simple form $Q_{\vartheta}=-\eta\dot{\vartheta}$ ($\eta>0$), 
we may write the equations of motion as
\begin{equation}
        \ddot{\vartheta}+\frac{\eta}{I}\dot{\vartheta}+
        \omega_{0}^{\,2}
        \left \{ \sin(\vartheta-r\phi)-
        \frac{1+r}{r}\sin[\vartheta-(1+r)\phi] \right \} =0,
        \label{eq:rod1}
\end{equation}
where
\begin{equation}
        \omega_{0}^{\,2}=r(1+r)\frac{mLR}{2I}\omega_{d}^{\,2}.
\end{equation}
If we introduce now a new variable $\theta=\vartheta-r\phi$ we 
can bring the equations of motion to the  arguably  simpler form:
\begin{equation}
        \ddot{\theta}+\frac{\eta}{I}\dot{\theta}+
        \omega_{0}^{\,2}\sin\theta=
        -\frac{\eta}{I}\omega_{d}+
        \frac{1+r}{r}\omega_{0}^{\,2}\sin(\theta-\phi).
        \label{eq:rod2}
\end{equation} 
This equation is nonlinear with a 
coupled term, and must be solved numerically. Therefore, for 
convenience we will write it in a dimensionless form by
introducing the dimensionless time variable $\tau=\omega_{0}t$.
Equation (\ref{eq:rod2}) then becomes
\begin{equation}
        \ddot{\theta}+\frac{1}{Q}\dot{\theta}+\sin\theta=
        -\frac{ar}{Q}+\frac{1+r}{r}\sin(\theta-\phi),
        \label{eq:rod3}
\end{equation}
where
\begin{eqnarray}
        a=\frac{\omega_{d}}{\omega_{0}}=\sqrt{\frac{2L}{3r(1+r)R}} 
        \nonumber \\
        Q=\frac{\omega_{d}}{\eta}\sqrt{r\frac{1+r}{2}mlRI},
\end{eqnarray}
and
\begin{equation}
        \phi=\omega_{d}t=a\tau.
\end{equation}
Since we assume throughout this paper that $\omega_D$ is constant,
the state of this system can be uniquely described in 3-dimensional 
phase space $(\theta,\dot{\theta},\phi)$. Furthermore, 
from the fact that $\phi$ depends linearly on time we may infer 
that trajectories in phase space will be ``smooth'' in the $\phi$ direction, 
and that all 
chaotic features, if any, can be observed in the Poincar\'{e} section 
$\phi=$ constant. In this way the problem of analyzing this three dimensional 
dynamical flow is reduced to the analysis of a two dimensional map.

Equation (\ref{eq:rod3}) has a similar form to that of the equation of
motion for the driven pendulum, as well as to the
Mathieu equation; however, there
are some essential differences. The virtual drive
frequency $a$ in the normalized equation (\ref{eq:rod3}) does not depend upon
the drive  frequency $\omega_{d}$ -- the  angular velocity of the moving
gear -- as it is constant which depends only on the 
dimensions of the rod and
gears. On the other hand, the ``quality'' factor $Q$ is proportional to
$\omega_{d}$, so a change of drive frequency will eventually appear as a
change of dissipation in the system.

\section{Routes to Chaos} % Occurrence of Chaos
\label{p2}

\begin{figure}%[H]
\centerline{\epsfig{file=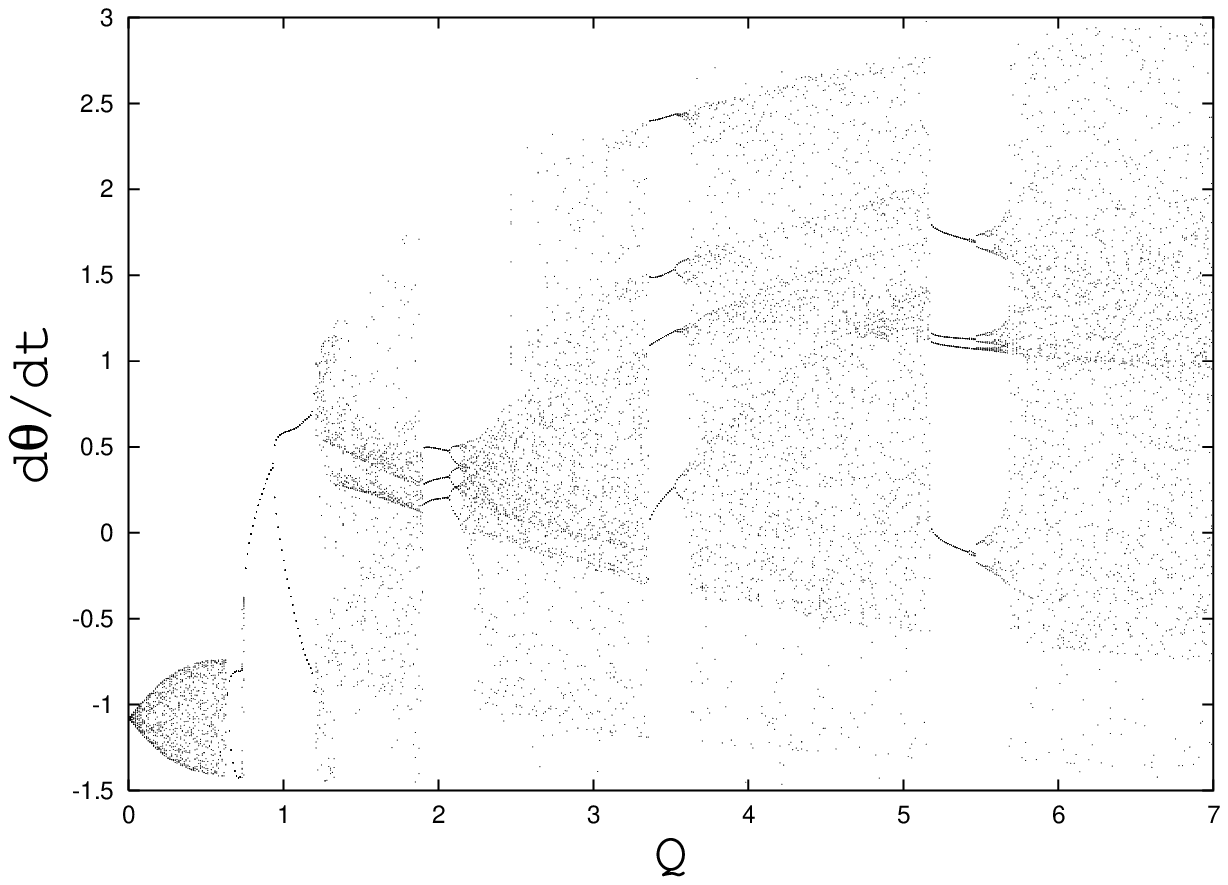, width=7cm}}
\centerline{\epsfig{file=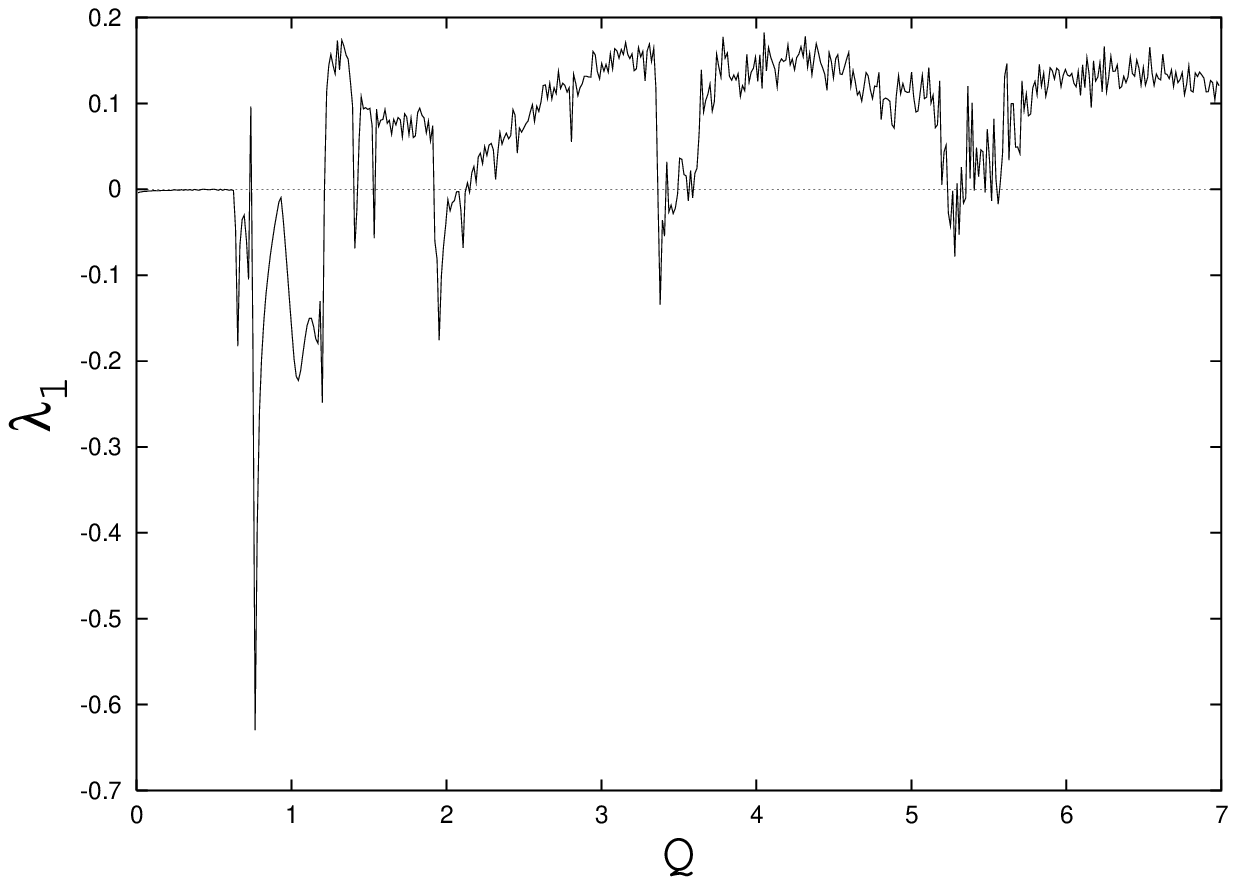, width=7cm}}
        \caption{(a) Bifurcation graph showing broad chaotic 
          region over segment of $Q$. 
          $a=1.2$, $r=0.9$. (b) Corresponding change of the largest 
          Lyapunov exponent.}
        \label{fig:chaos}
\end{figure}

For our system, we observe three characteristic regions 
in the parameter space -- quasiperiodic, periodic and chaotic. 
In order to locate these regions we investigate 
the parameter space using bifurcation diagrams. Such diagrams 
show the evolution of the attractor for a dissipative 
system with a change of the system's parameters. We choose to plot 
a projection of the Poincar\'{e} section $\phi=0$ 
on the $\dot \theta$ 
axis against the system's parameter $Q$, while keeping other two parameters 
fixed. We repeat this procedure for various values of $a$ and $r$.

With the increase of $Q$ from zero on, the motion of the system is 
quasiperiodic, which is observed as a smeared region in the bifurcation graph. 
For certain value of $Q$ the motion suddenly becomes periodic, and with 
a further increase of the quality factor $Q$ the
system reaches chaos through a sequence 
of period doublings. For the parameters $a,r<0.5$, the 
quasiperiodic region is relatively short, and chaos is found only in a narrow 
region of $Q$ values. 
The richest chaotic structure is found for the choice of parameters 
$a,r \sim 1$. Finally, for $a,r > 1$ quasiperiodic motion becomes 
dominant over the 
range of $Q$, and periodic and possible chaotic behavior are found only 
in narrow isolated regions.

\begin{figure}%[H]
\centerline{
  \epsfig{file=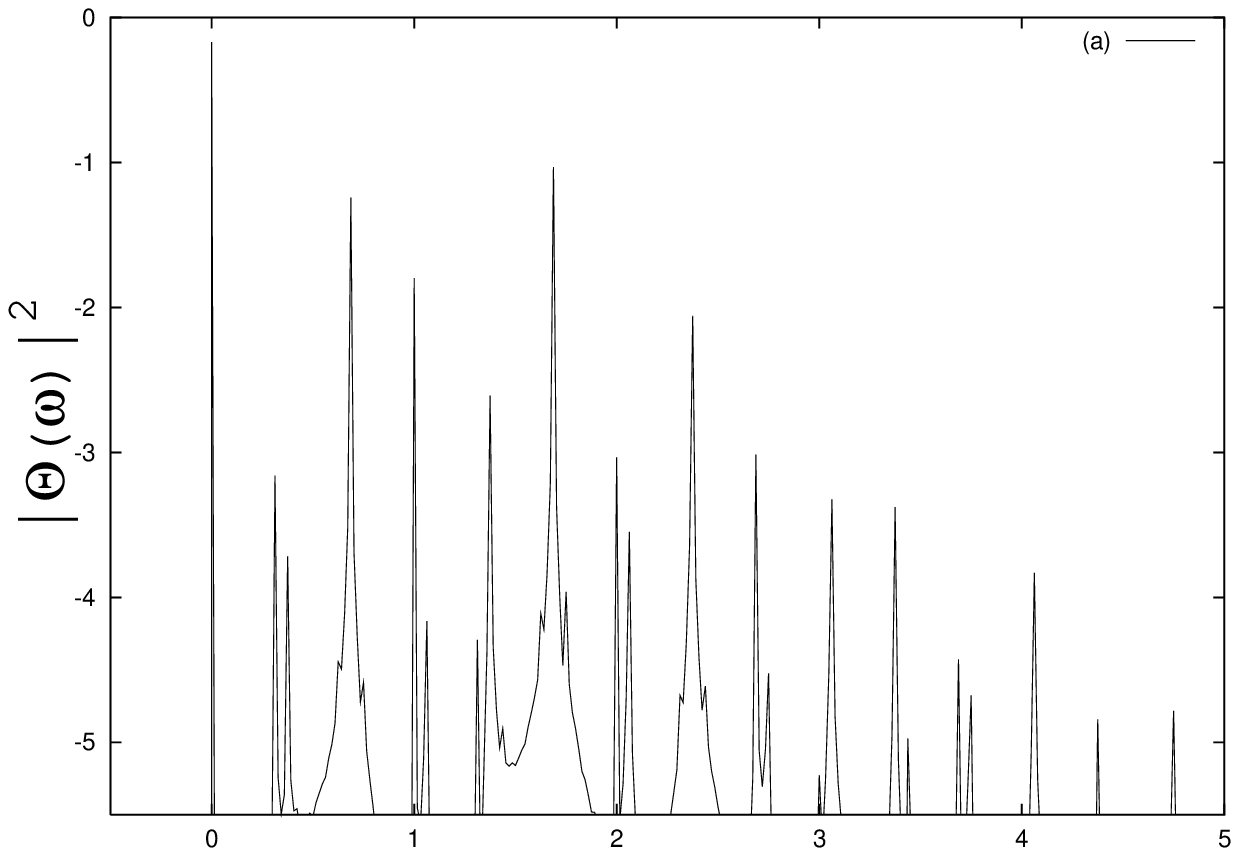, height=5cm}
  \epsfig{file=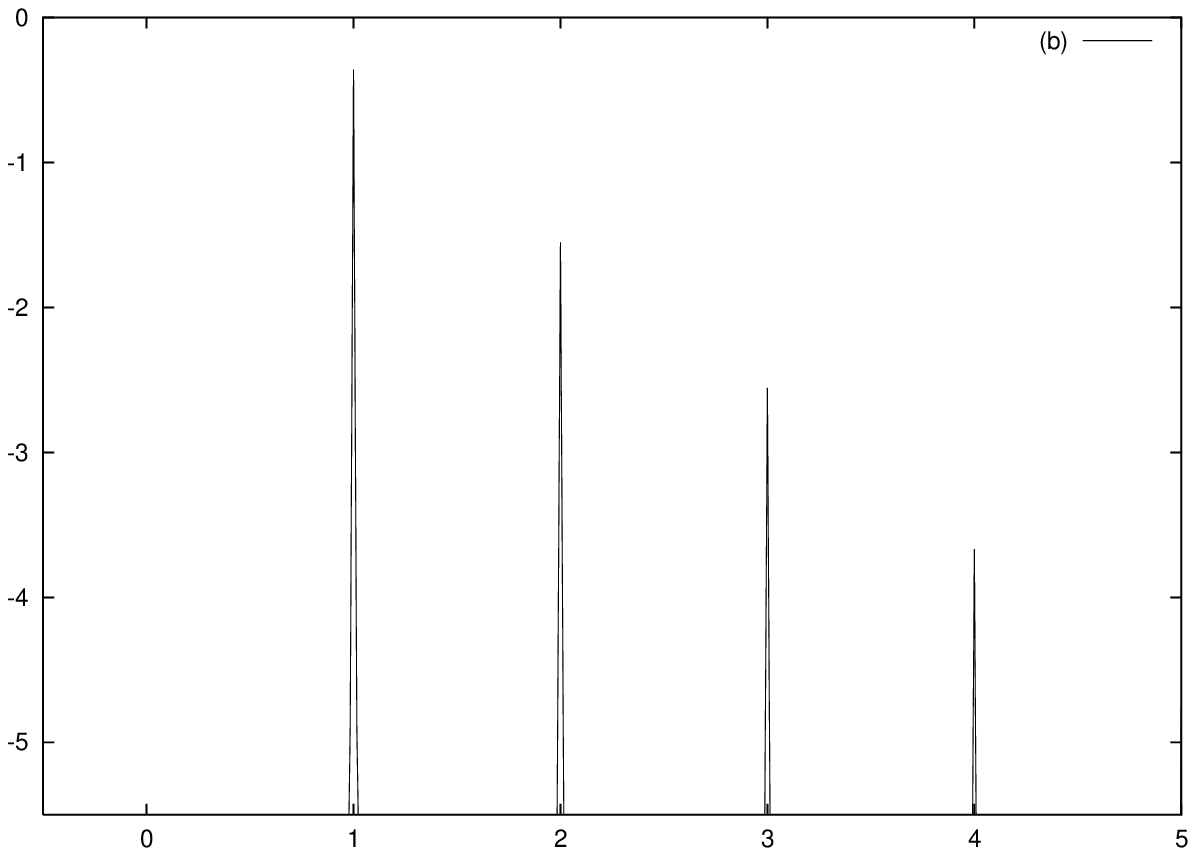, height=5cm}       
}
\centerline{
  \epsfig{file=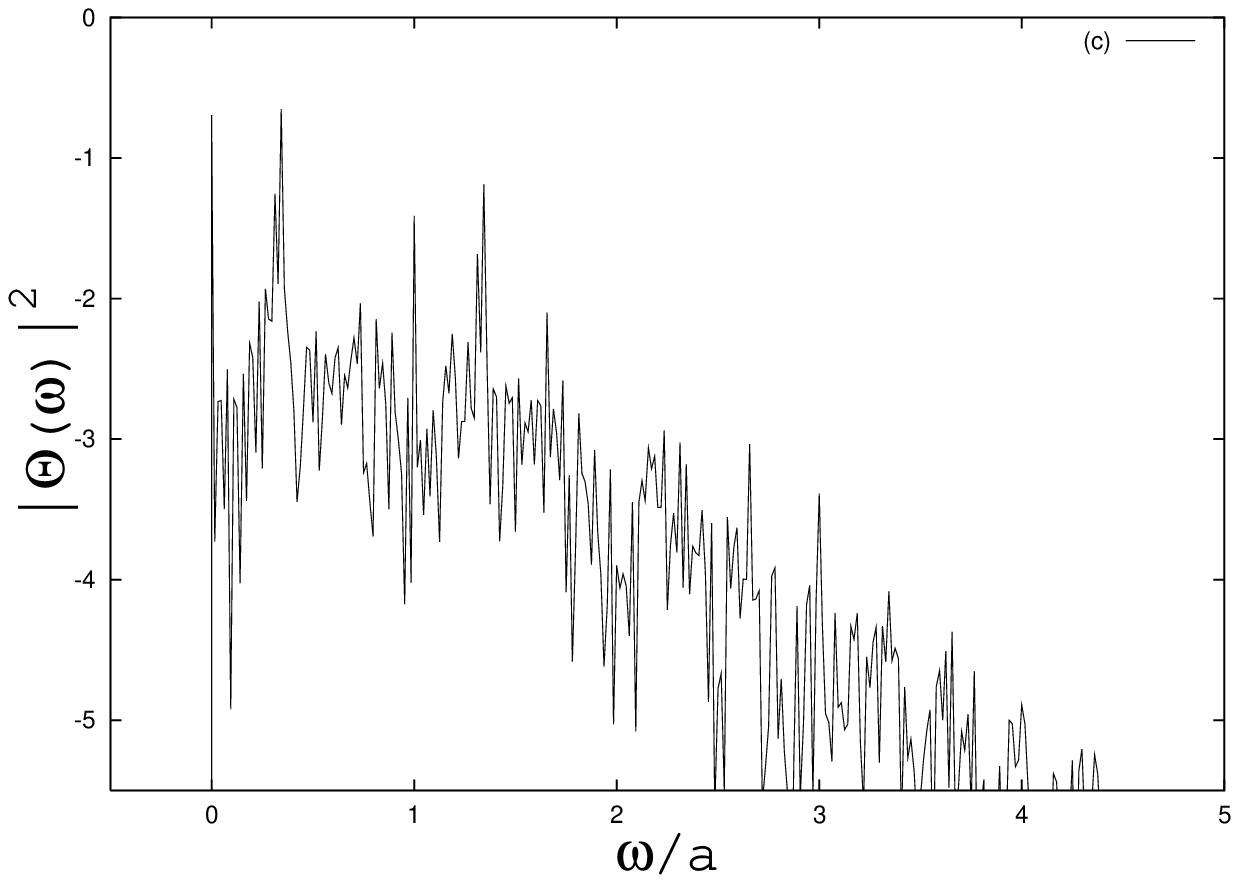, height=5cm}
  \epsfig{file=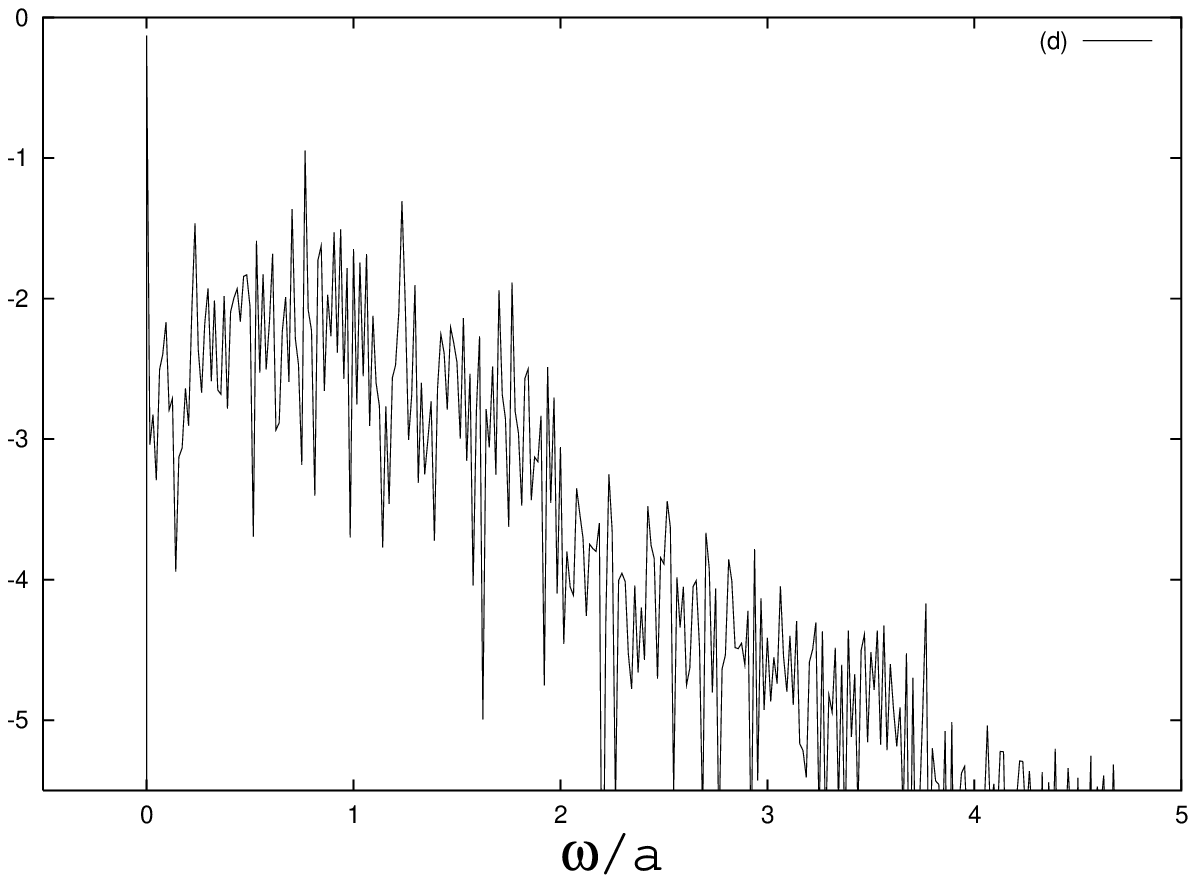, height=5cm}
}
        \caption{ Power spectra for different values of quality factor $Q$: 
        (a) $Q=0.2$ motion is quasiperiodic, 
        (b) $Q=0.9$, motion is periodic with period one, (c) $Q=1.8$ and 
        (d) $Q=6.5$, motion is chaotic. 
        $\Theta(\omega)$ is the Fourier transform of $\dot{\theta}(t)$. 
        The frequency on the $x$--axis is expressed in terms of 
        the virtual drive frequency $a$.}
        \label{fig:ffts}
\end{figure}

Solid proof of chaos is the existence of at least one positive 
Lyapunov exponent. The spectrum of Lyapunov exponents gives us, besides
qualitative information about a system's behavior, also a quantitative measure 
of the system's stability \cite{Ose68,Ben80,Shi79,Wol84}. 
Evolution of the Lyapunov exponent spectrum with a
change of parameters gives us additional information about the system's 
behavior to that from bifurcation graphs. 
Combined, these two methods are reliable tools for examining a system's 
behavior and routes to chaos. Figure \ref{fig:chaos} gives a comparison 
of the
bifurcation diagram and Lyapunov spectrum evolution for a range of $Q$.
Since the system is represented in 3 dimensional space it is sufficient to 
determine only the largest Lyapunov exponent. The other two can be inferred 
from Haken's theorem \cite{Hak83}, and the fact that the dissipation of 
the system 
is constant throughout phase space, so the sum of Lyapunov 
exponents must equal to its numerical value $\sum_{i}\lambda_{i}=1/Q$. 

\begin{figure}%[H]
        \label{fig:pcr}

\centerline{\epsfig{file=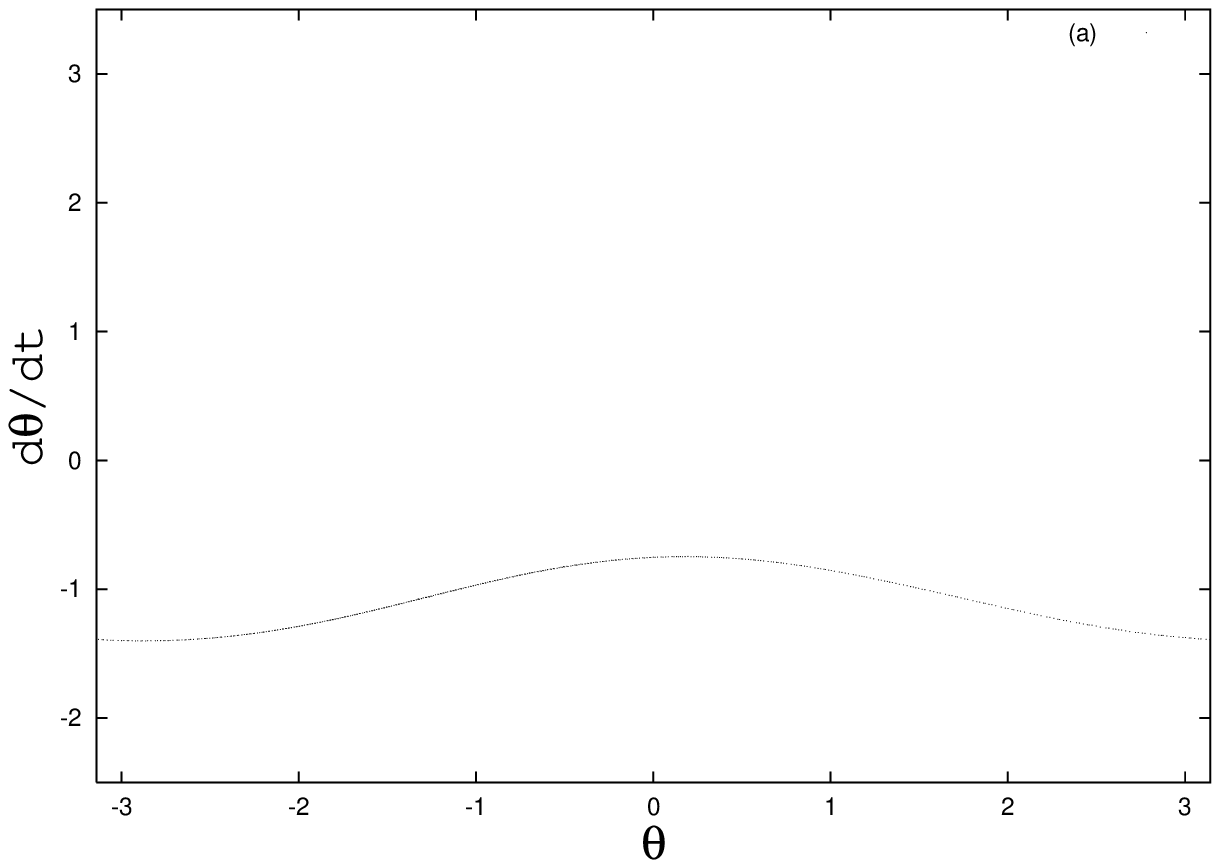, width=7cm}}
\centerline{\epsfig{file=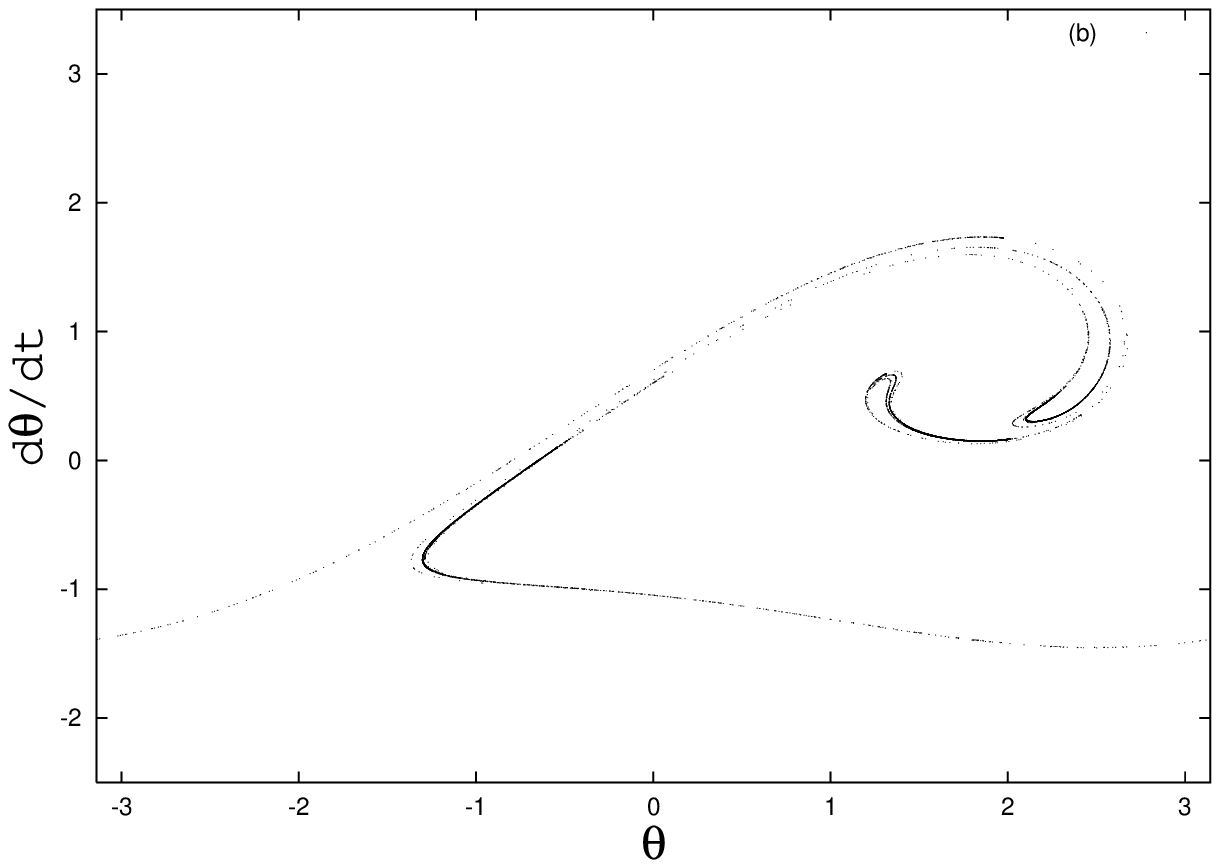, width=7cm}}
\centerline{\epsfig{file=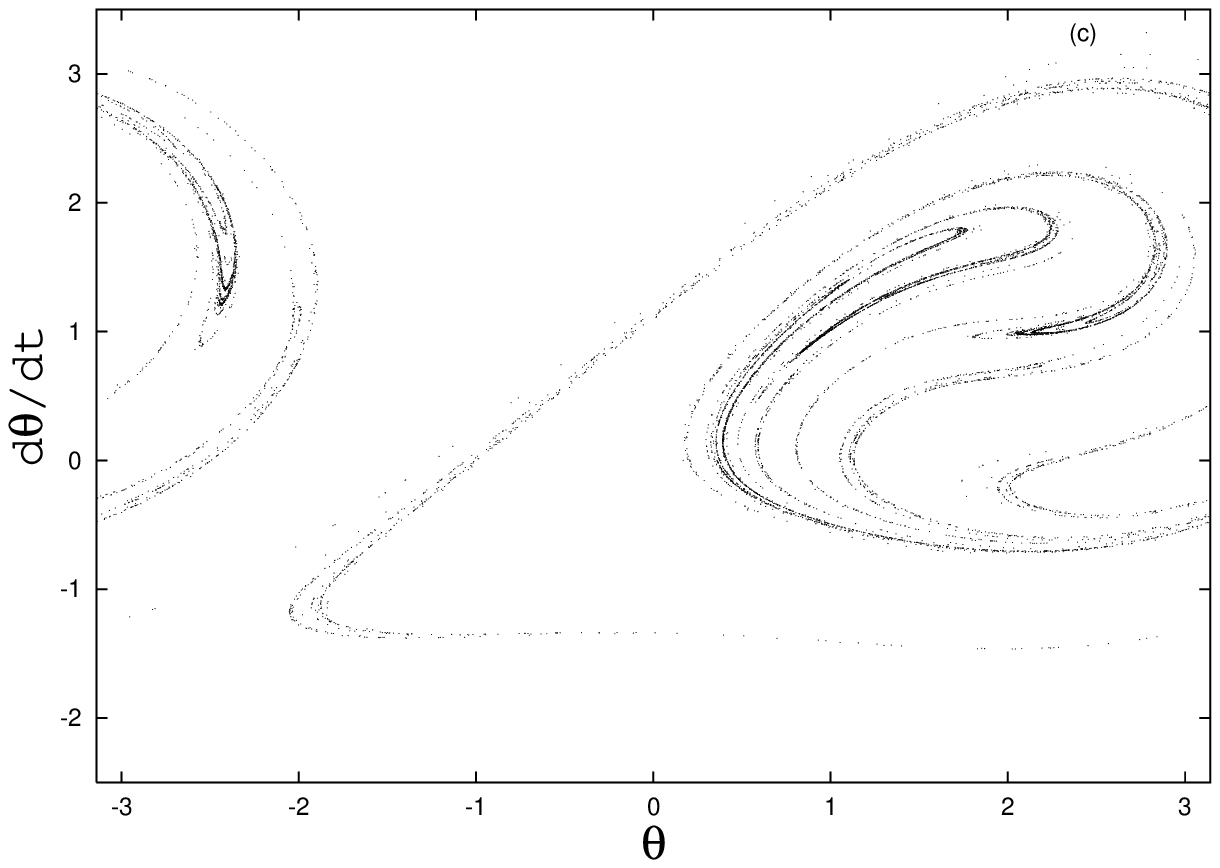, width=7cm}}
\caption{ (a) Poincar\'{e} section of the quasiperiodic attractor is 
a smooth line. $a=1.2$, $r=0.9$, $Q=0.5$. (b) Poincar\'{e} section of
the chaotic attractor reveals a kneading sequence. $Q=1.8$. (c) Higher 
dimensional chaotic attractor. $Q=6.5$.}

\end{figure}

A detailed description of the parameter space for this system is 
given in Ref.\cite{Pel00}. Here we illustrate its main features with 
a bifurcation diagram 
plotted for parameters $a=1.2$ and $0.9$ (Fig.~\ref{fig:chaos}a). The
smeared region in the bifurcation diagram that is observed for the values 
$0<Q<0.6378 \ldots$ corresponds to a quasiperiodic 
behavior of the system. Although it is generally possible to notice 
differences between 
chaotic and quasiperiodic regions in a bifurcation graph, it is necessary 
to employ other methods in order to have solid evidence of a system's 
behavior.

Analysis of the power spectrum of the numerical solution is one of the methods 
that is particularly suitable for examining quasiperiodic motion. A
power spectrum for $Q=0.5$ (Fig.~\ref{fig:ffts}a) shows that the
motion of the system is a superposition of a
finite number of harmonic modes, and therefore is not chaotic. The motion is 
not periodic though, since frequencies of harmonics are incommensurate, and 
the trajectory in phase space never retracts itself. 
The attractor for the system is a smooth 2-dimensional surface 
in phase space. The Poincar\'{e} section 
for $Q=0.5$ (Fig.~\ref{fig:pcr}a) indeed reveals smooth attractor.

For values of 
$0.7442 \ldots <Q<0.9300 \ldots$ it is clear from the bifurcation graph that the
system's behavior is periodic with period that is equal to the
virtual drive period $2 \pi / a$. This is also seen in the power 
spectrum for $Q=0.9$ in Figure ~\ref{fig:ffts}b. 
With an increase of $Q$ over 
$1.2040 \ldots$ the system becomes chaotic through a period doubling sequence. 
The smeared region in the 
bifurcation diagram indicates that the attractor for the system has a complex 
structure in that region of the parameter space. The power spectrum also 
shows complexity of the system's behavior and indicates chaos in 
that region of parameter space. The power spectrum does not consist 
of a finite number of harmonics and their integer multiples any more, but 
is rather a broad band over the frequency axis 
(Figs.~\ref{fig:ffts}c and ~\ref{fig:ffts}d). 
In order to obtain a solid evidence of chaos one 
has to show that at least one Lyapunov exponent for the system is positive 
in that region \cite{All97}. Indeed, Figure ~\ref{fig:chaos}b shows 
that the largest 
Lyapunov exponent is positive in this region of parameter space, apart 
from windows of periodic behavior.

In particular, at $Q=1.8$ and $Q=6.5$ the largest Lyapunov exponents are 
found to be 
$\lambda_{1}=0.093 \pm 0.005$ and $\lambda_{1}=0.106 \pm 0.005$, respectively. 
The system's attractors for these values of parameters 
(Figs.~\ref{fig:pcr}b and ~\ref{fig:pcr}c) 
show characteristic stretching and folding pattern, indicating their fractal 
structure. 

In the quasiperiodic region the attractor has dimension 2, as the motion is
performed over a smooth surface, while the
limit cycles have dimension 1. Since
motion takes place in 3-dimensional phase space it is reasonable to expect 
that strange attractors have dimension between 2 and 3. We can use the fact
that the attractor is always smooth in the $\phi$ direction, so in order to
estimate its dimension we can estimate the dimension of its 
Poincar\'{e} section
and obtain the attractor's dimension simply by adding one to it. 

\begin{table}%[H]
\begin{center}
\begin{tabular}{|c|c|c|c|c|}
\hline 
%  & & & & \\
  $~~Q~~$ &  $D_{L}$ &  $D_{0}$ &  $D_{1}$ & $D_{2}$ \\
\hline
% & & & & \\
$0.5$  & $2.000 \pm 0.001$ & $2.002 \pm 0.003$ & $2.007 \pm 0.006$ &
        $1.999 \pm 0.002$ \\
% & & & & \\
$1.8$  & $2.14 \pm 0.01$ & $2.11 \pm 0.01$ & $2.044 \pm 0.007$ &
        $2.012 \pm 0.003$ \\
% & & & & \\
$6.5$  & $2.41 \pm 0.01$ & $2.39 \pm 0.01$ & $2.389 \pm 0.004$ &
        $2.299 \pm 0.001$ \\
\hline
\end{tabular}
\caption{Various dimensions of the attractors.}\label{tab:dim}
\end{center}
\end{table}

In order to estimate the dimension of an attractor lying in $n$--dimensional 
phase space, the attractor, first, has to be covered by a grid of 
$n$-dimensional hypercubes of size $\epsilon$, and then the probability of 
finding a point of the 
attractor in each hypercube $P_{i}$ has to be determined. 
The index $i$ here refers to 
a particular cube. In our case we consider a 2-dimensional Poincar\'{e} 
section, so the hypercubes are actually squares. A general expression 
for the dimensions of the $q^{\rm th}$-order is then given by:
\begin{equation}
  D_{q}=\lim_{\epsilon \rightarrow 0} \frac{1}{q-1}
  \frac{\ln \sum_{i}P_{i}^{\,q}}{\ln \epsilon},
\end{equation}
where the summation is over all hypercubes where $P_{i}>0$ \cite{Gra84}. 
The parameter $q$ ranges from 
$-\infty < q < \infty$, and for $q_{1}>q_{2}$ we have 
$D_{q_{1}} \le D_{q_{2}}$. 
The most commonly used dimensions are $D_{0}$, $D_{1}$, and $D_{2}$, or 
the capacity, information, and correlation dimension, respectively. 
We will limit ourselves to estimating these three dimensions
only.

On the other hand, the Lyapunov dimension of the attractor is defined as:
\begin{equation}
        D_{L}=j+\frac{\lambda_{1}+\lambda_{2}+...+\lambda_{j}}{|\lambda_{j+1}|}
\end{equation}
where $j$ is a largest integer for which 
$\lambda_{1}+\lambda_{2}+\ldots+\lambda_{j} \geq 0$.
Kaplan and Yorke conjectured that the
Lyapunov dimension represents an upper limit to the information
dimension, i.e. $D_{L} \ge D_{1}$ \cite{Kap79,Led81}. It is to be
expected though that numerical values of these four dimensions fall
pretty close to each other \cite{Far83}.

We test this conjecture on our calculations for the quasiperiodic 
attractor (Fig.~\ref{fig:pcr}), and as
expected we obtain the
Lyapunov, capacity, information, and correlation dimensions
to be equal 2, within the limits of
uncertainty of our numerical methods. 
Furthermore, we can estimate these dimensions for the
attractors in Figures ~\ref{fig:pcr}b and ~\ref{fig:pcr}c. 
Both chaotic attractors have
noninteger dimensions (Table \ref{tab:dim}),
 confirming their fractal structure. In all
three cases the inequality $D_L \ge D_1$ holds, suggesting that this
system conforms Kaplan-Yorke conjecture.

\section{Coexisting Attractors and Crisis}\label{sec:cox}

\begin{figure}[H]
\centerline{\epsfig{file=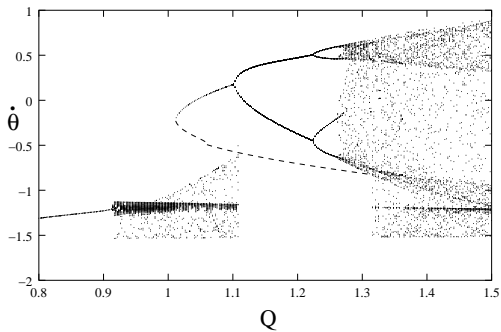, width=7cm}}
        \caption{Bifurcation diagram indicating coexisting
        attractors. Dashed line shows unstable limit cycle.}
\label{fig:lock} 
\end{figure}
\begin{figure}[h]
\begin{center}
\leavevmode
\epsfxsize=7 cm
        \epsfbox{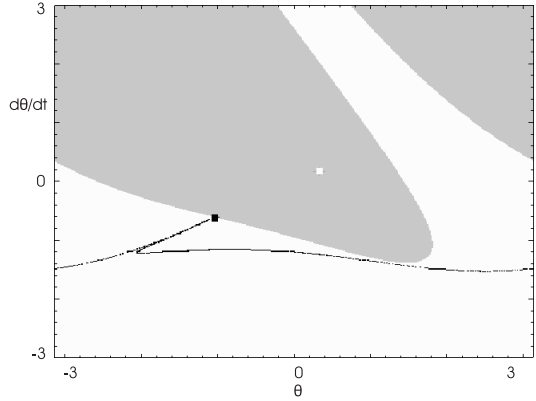}
        \caption{ Poincar\'{e} section of the chaotic attractor at the verge
        of crisis, shown within its basin of attraction in. White dots
        represent a stable, and black dots an unstable, limit cycle. $Q=1.8$.}
        \label{fig:bas110}
\end{center}
\end{figure}

An interesting feature of this system is that it often has 
periodic and chaotic attractors coexisting at the same point 
in parameter space. This means that the stability of 
this system may depend on the initial conditions -- whether we 
choose it in the basin of periodic or chaotic attractor. 
Here we shall present two examples which illustrate the importance of 
understanding system's dynamics in order to properly assess its 
stability. 

The bifurcation diagram \ref{fig:lock} shows that 
a tangent bifurcation \cite{Man79} 
occurs $Q=1.0105 \ldots$, which creates a pair of stable and
unstable limit cycles in the phase space, while a chaotic attractor still 
exists. Accordingly, the phase space splits 
into a basin of the stable limit cycle and a basin of the
chaotic attractor. The boundary between the two basins appears
 to be a smooth curve (Fig.~\ref{fig:bas110}). 
With an increase of the
control parameter $Q$ the basin of attraction for the limit cycle
expands until the unstable limit cycle, which lies on its boundary,
collides with the chaotic attractor. The chaotic attractor
experiences a boundary crisis \cite{Gre83} and disappears along with
its own basin. The basin of the limit cycle then suddenly expands,
occupying the whole phase space.
With a further increase of the control parameter the limit cycle
evolves into a chaotic attractor through a sequence of period doublings.
The newly created chaotic
attractor expands in size as the parameter $Q$ increases, and
eventually collides with the unstable limit cycle. The system comes to a
crisis again, but since the unstable orbit is not at the basin boundary any
more the crisis is internal, and the attractor suddenly expands its size
\cite{Gre83}. This expanded attractor contains features of the
chaotic attractor destroyed
in the boundary crisis, indicating that information about the system's
behavior at lower values of $Q$ has ``survived'' the crisis.

\begin{figure}[H]
\centerline{\epsfig{file=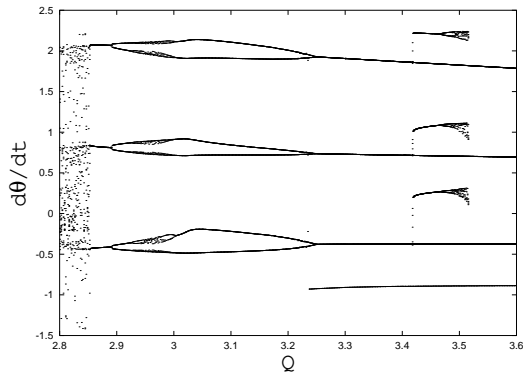, width=7cm}}
        \caption{ Bifurcation diagram shows creation of coexisting
        attractors. } \label{fig:basbif}
\end{figure}

\begin{figure}[H]
\begin{center}
\leavevmode
\epsfxsize=7 cm
        \epsfbox{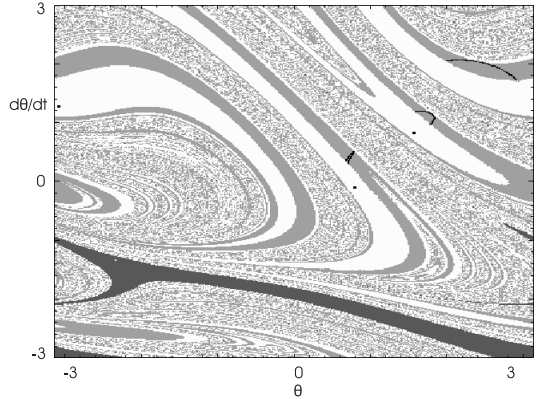}
        \caption{ Three coexisting attractors. Period-three limit cycle
          (black points) and period-one limit cycle (white point) are
          enlarged for clarity. $Q=3.515$.}
        \label{fig:bas3515}
\end{center}
\end{figure}

As we increase the control parameter $Q$ further,
we observe another occurrence of coexisting attractors
(Fig.~\ref{fig:basbif}). Tangent bifurcations at 
$Q=3.2398 \ldots$ and $Q=3.4177 \ldots$ create stable--unstable pairs 
of period--one and period--three orbits, respectively, so we have 
three coexisting attractors. Basin boundaries are now 
locally disconnected fractal curves \cite{McD85a}, as 
shown in Figure ~\ref{fig:bas3515}. Basins of attraction 
are highly interwoven in this region, and for practical purposes 
it is difficult to determine initial conditions that would lead to
a particular attractor. The newly created period--3 orbit quickly evolves to
a chaotic attractor with an increase of the control parameter
(Fig.~\ref{fig:basbif}), and disappears in a boundary crisis. 
Basins of attractions in Figures \ref{fig:bas110} and \ref{fig:bas3515} 
are plotted using software package {\em Dynamics 2} \cite{Nus}.

% Symbolic dynamics

\section{Partitioning of the Poincar\'e section}
\label{sec2}

We now turn to a symbolic analysis of this model. For this,
we shall choose convenient phase space $x=\theta /2\pi$, $y=\dot x$, 
and write the equations of motion in their nonautonomous form as
\begin{equation}
\label{e1}
\begin{array}{l}
\dot x = y,\\
\dot y = - y/Q + [-ar/Q-\sin(2\pi x) 
        + (1 + r)/r \sin(2\pi x - a \tau)]/(2\pi).\\
\end{array}
\end{equation}
To show its
richer properties in the phase and parameter spaces we fix $r = 1.088$
 and vary 
the values of $Q$ and $a$ in such a way that $a = 0.8 + 0.3Q$. 
All Poincar\'{e} sections are constructed in $\phi=27\pi /64$ plane 
of phase space. 
%All stroboscopic portraits are constructed by taking an initial
%phase
%$t_0 = 27\pi / (64a)$.

\begin{figure}
\centerline{
\epsfig{file=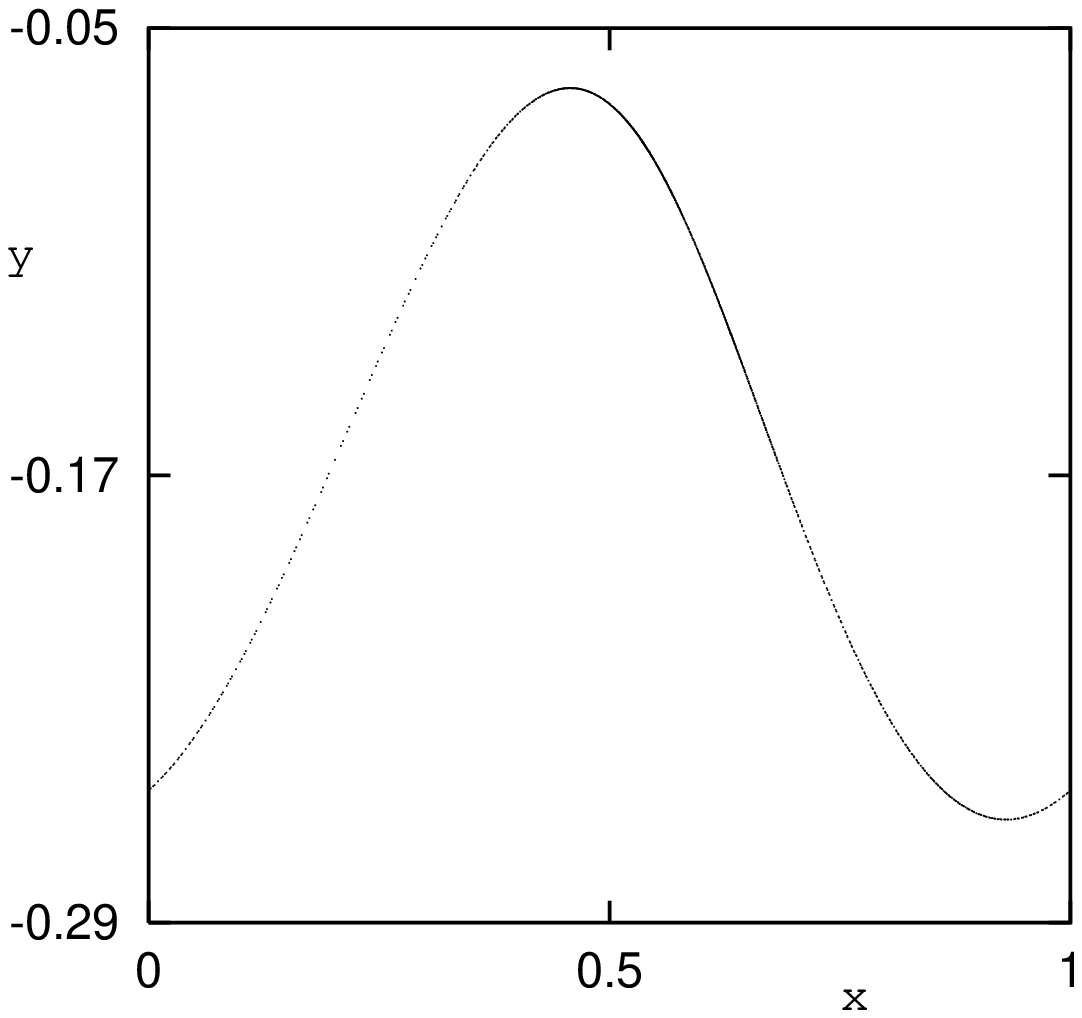,height=1.8in,width=1.6in}
\epsfig{file=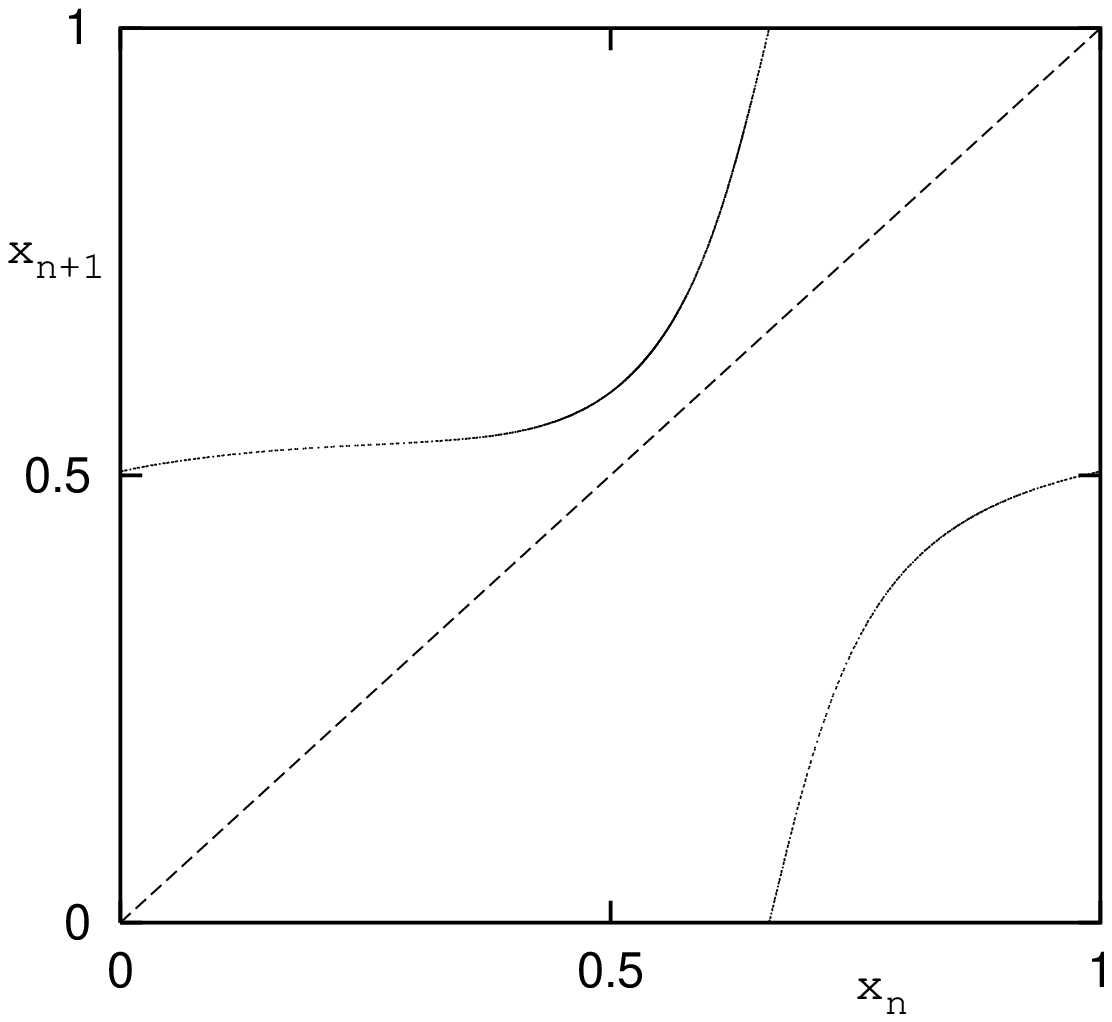,height=1.8in,width=1.6in}
\epsfig{file=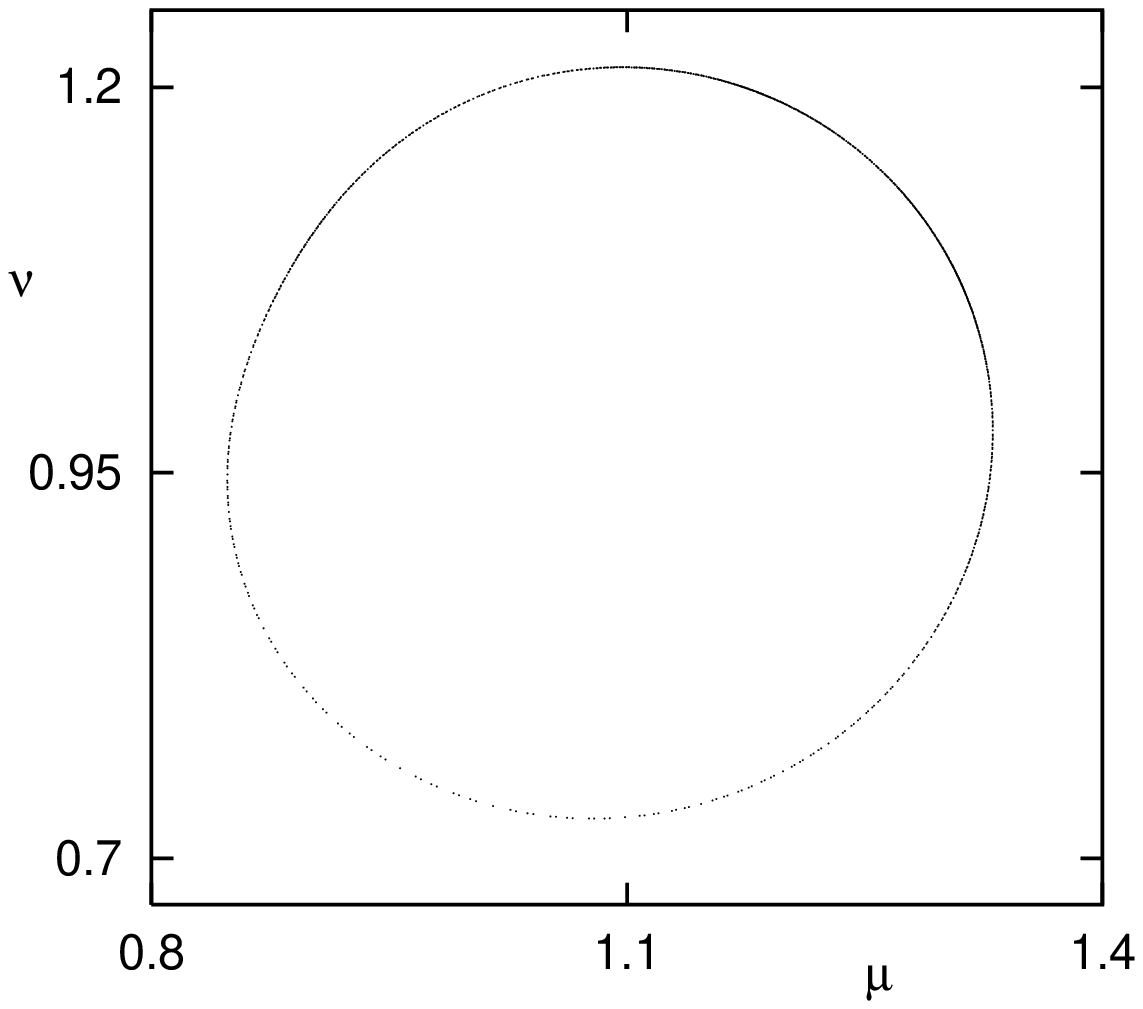,height=1.8in,width=1.6in}
}
\centerline{
\epsfig{file=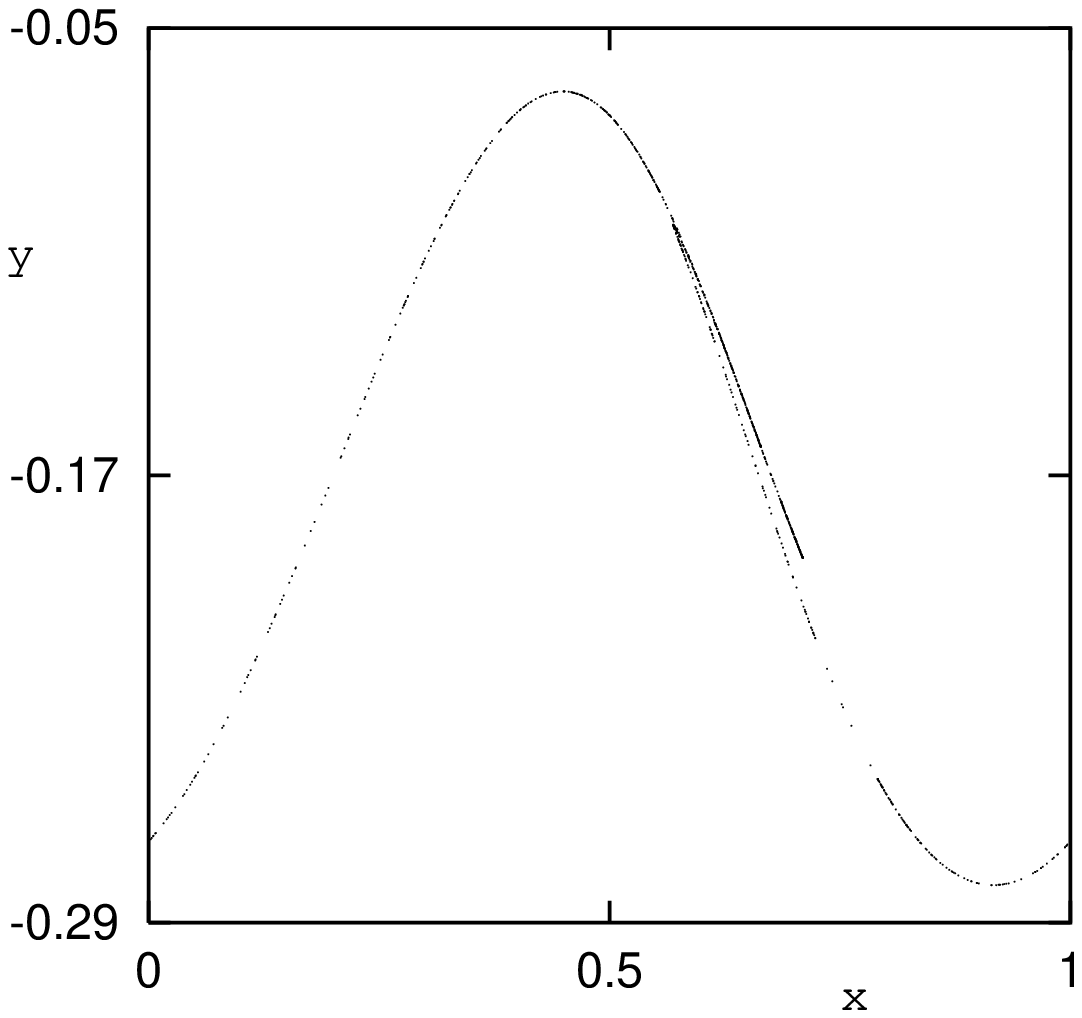,height=1.8in,width=1.6in}
\epsfig{file=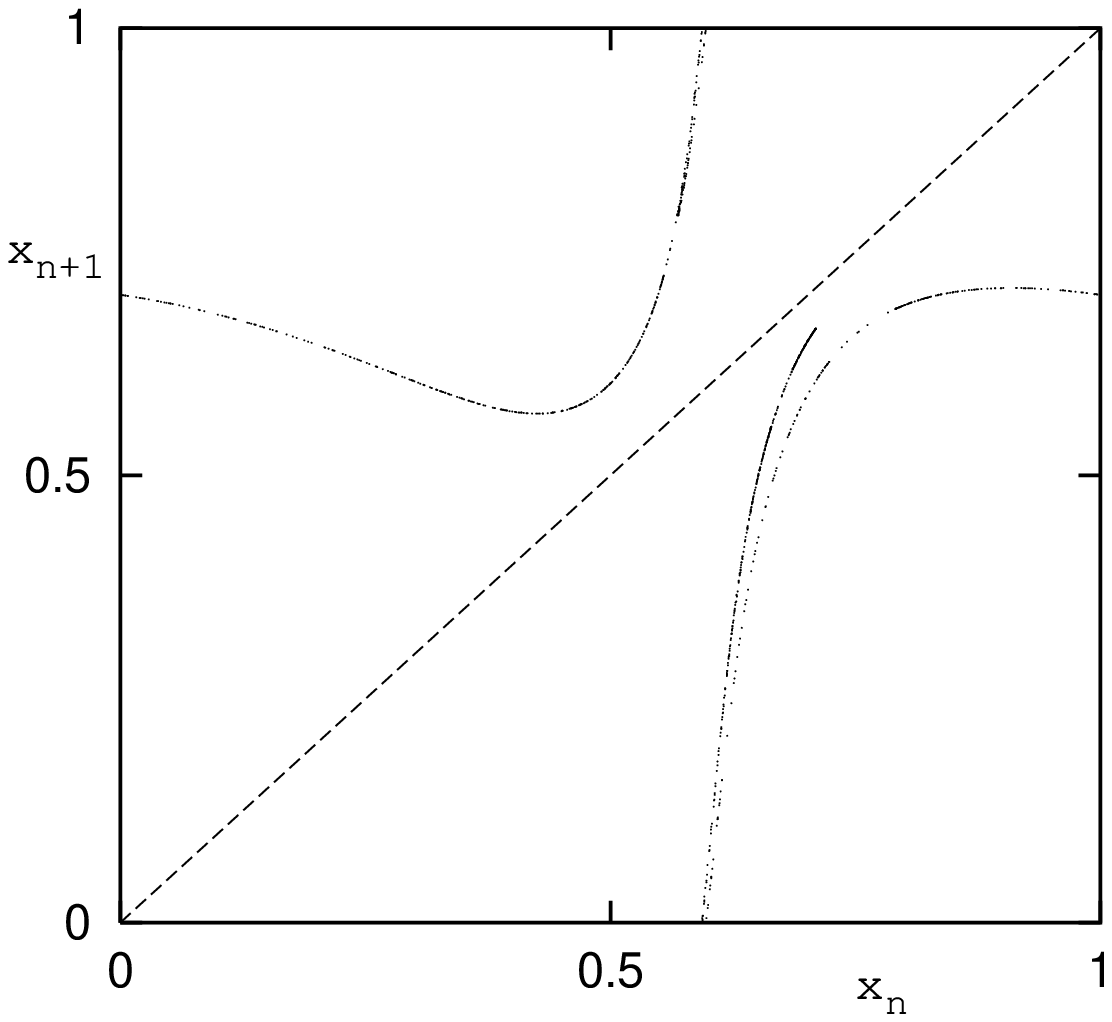,height=1.8in,width=1.6in}
\epsfig{file=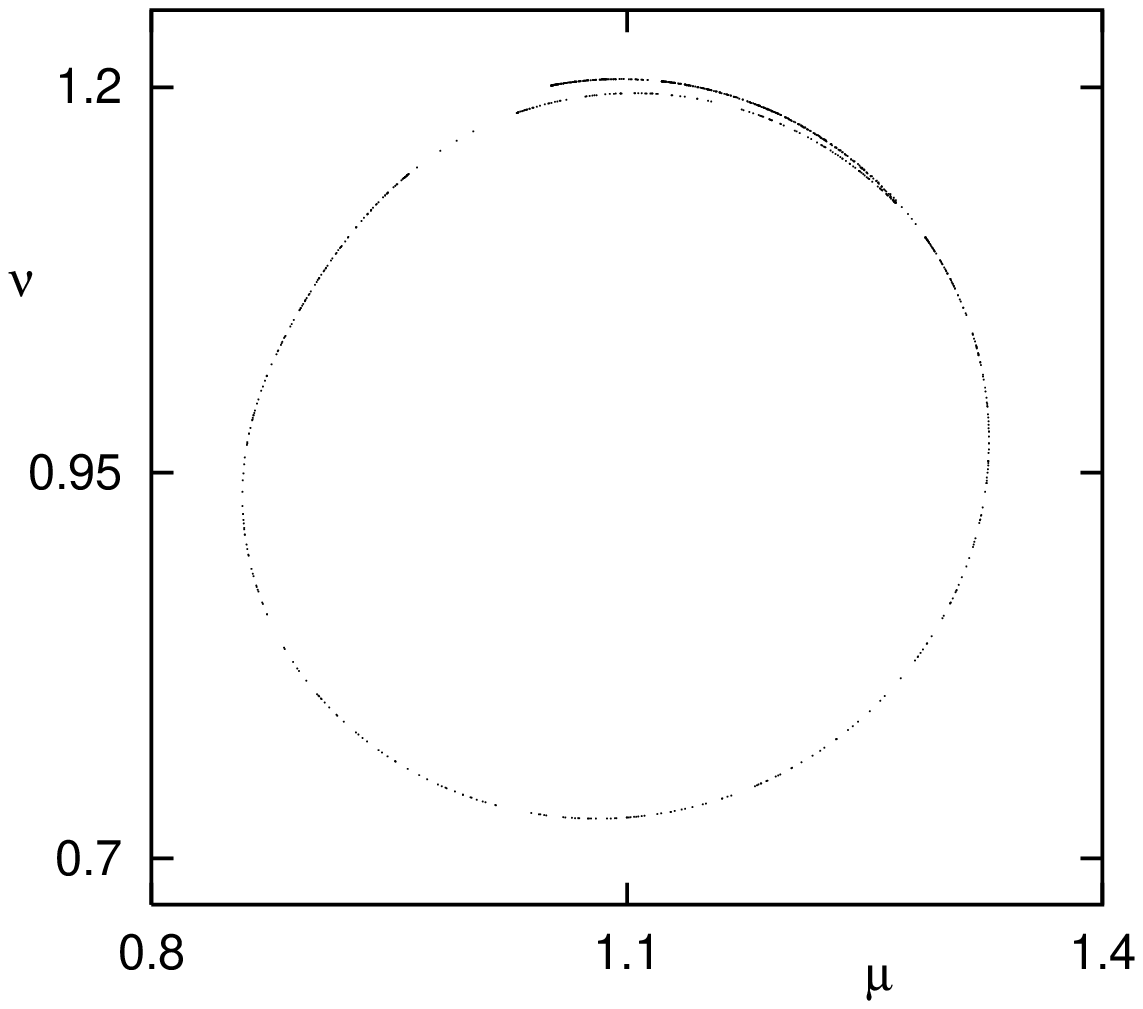,height=1.8in,width=1.6in}
}
\centerline{
\epsfig{file=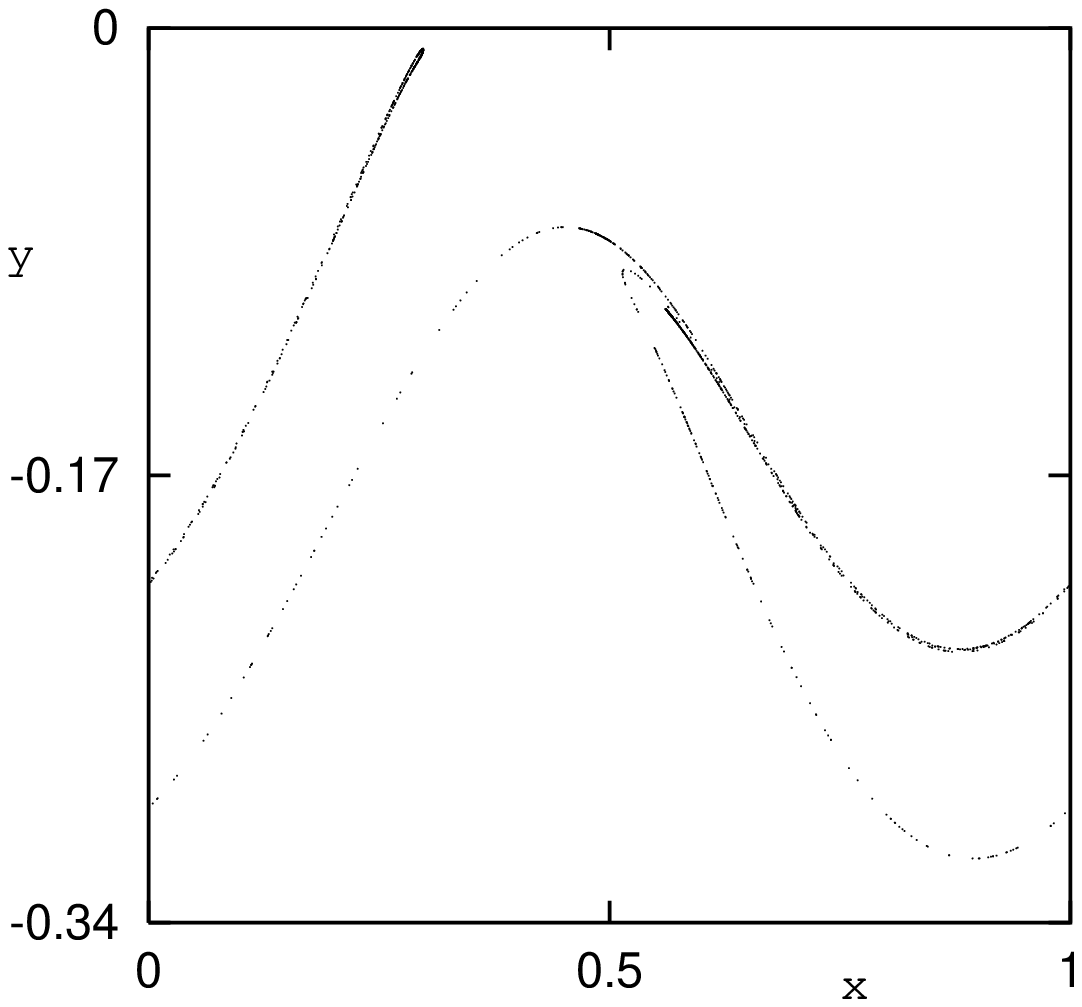,height=1.8in,width=1.6in}
\epsfig{file=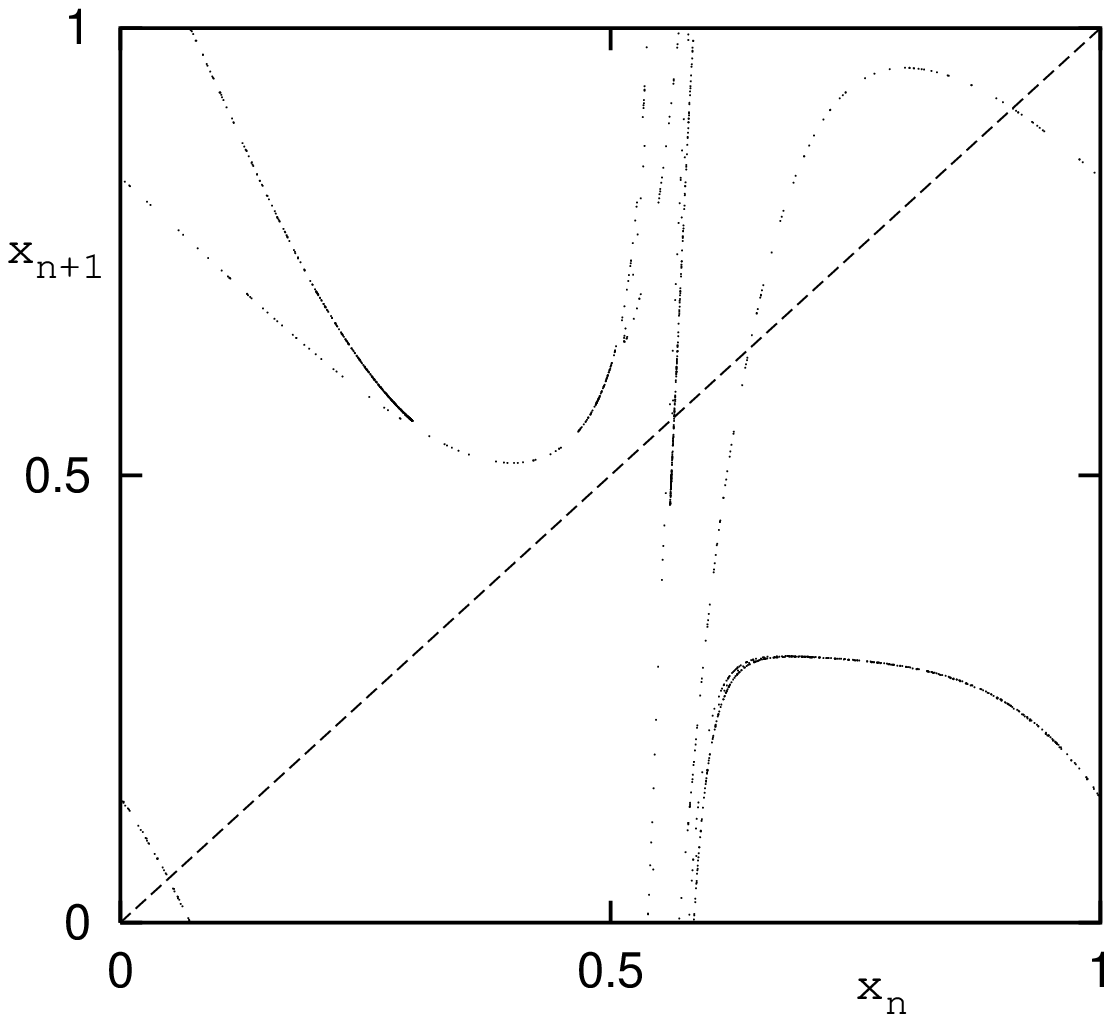,height=1.8in,width=1.6in}
\epsfig{file=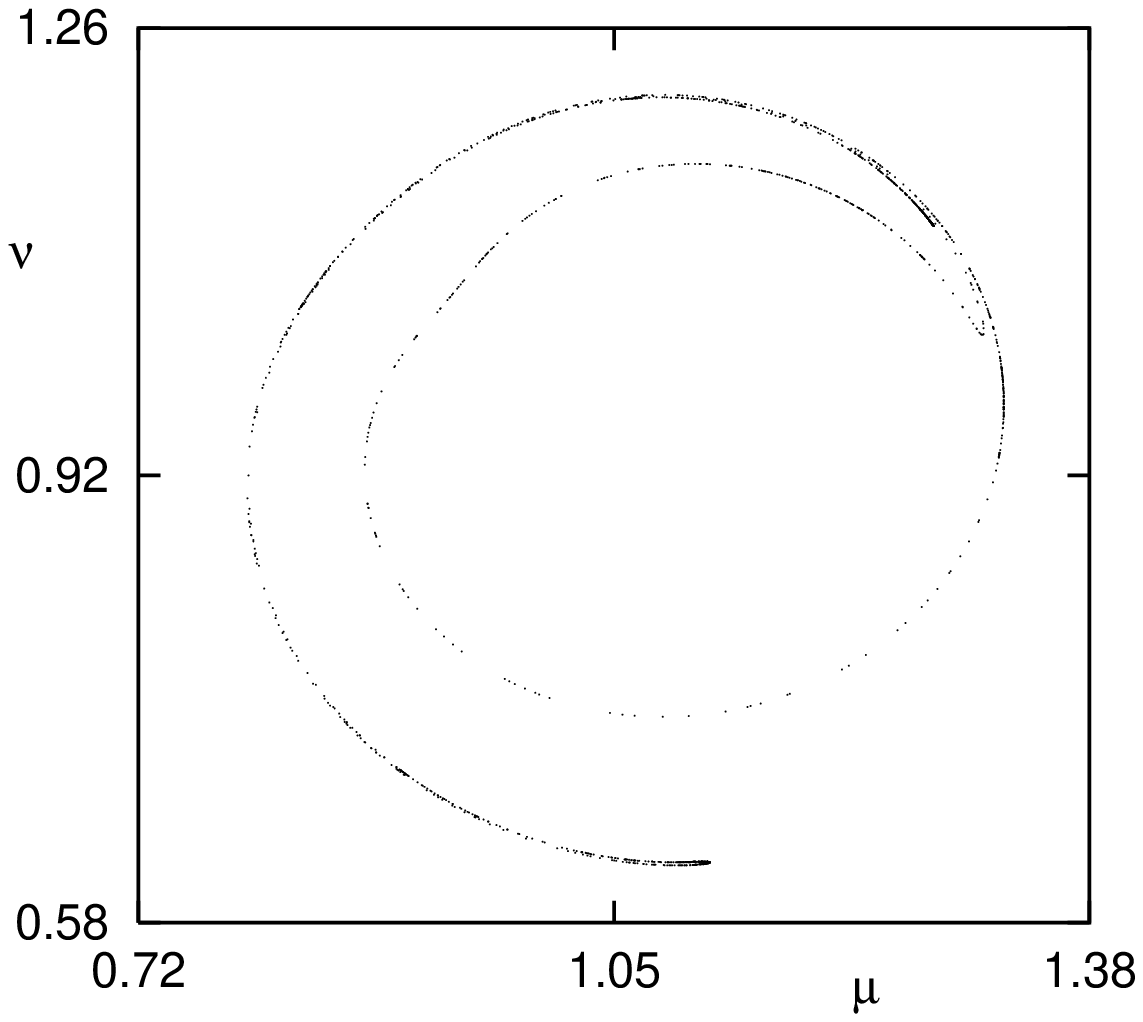,height=1.8in,width=1.6in}
}
\centerline{
\epsfig{file=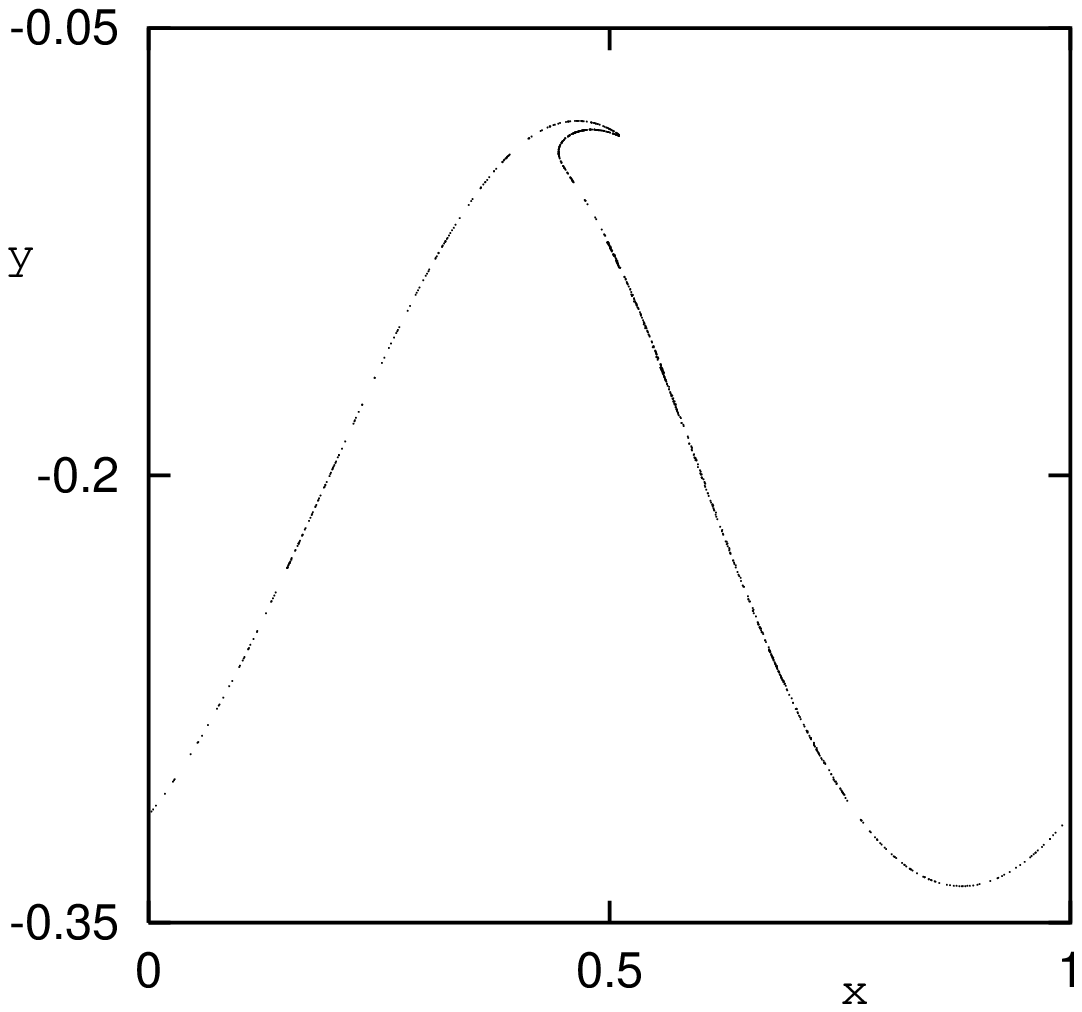,height=1.8in,width=1.6in} 
\epsfig{file=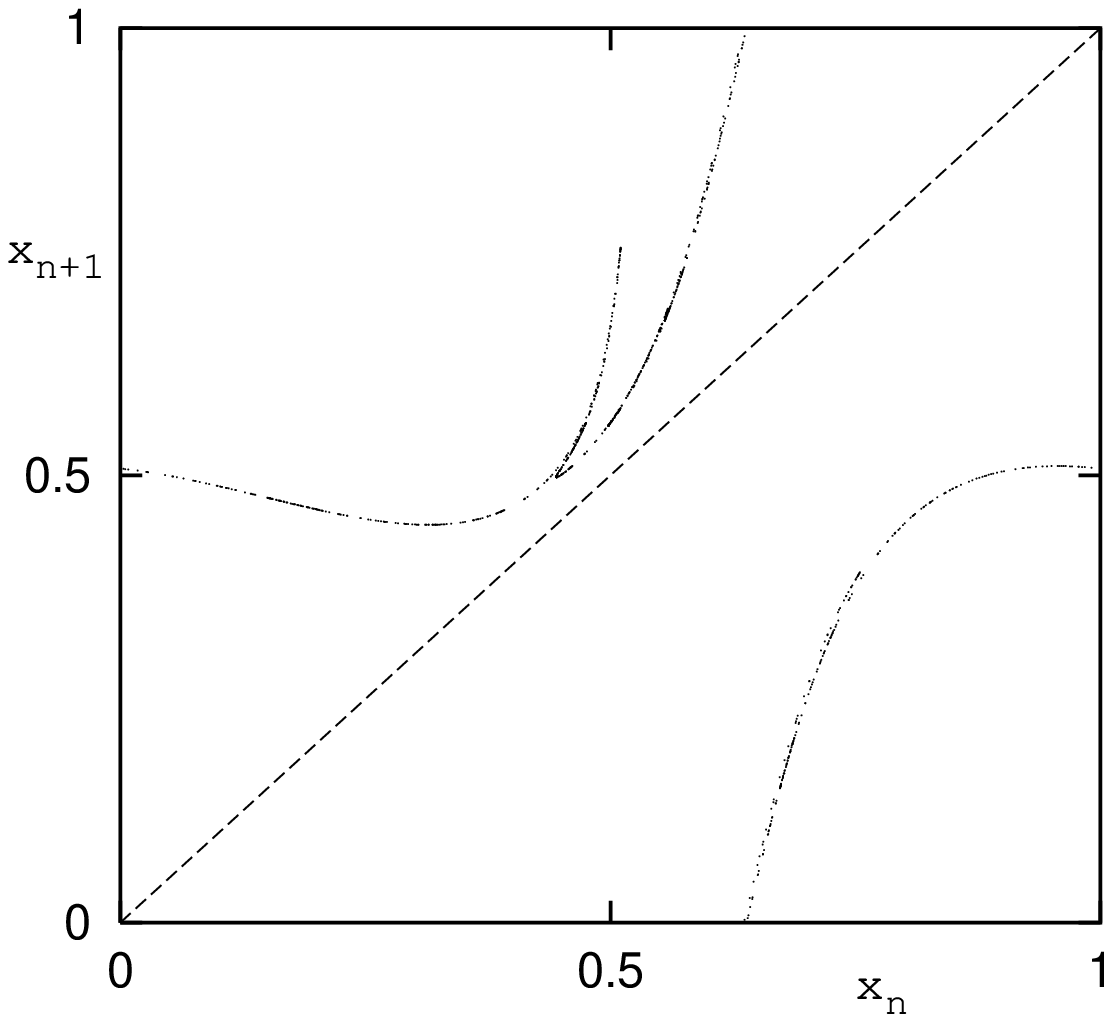,height=1.8in,width=1.6in}
\epsfig{file=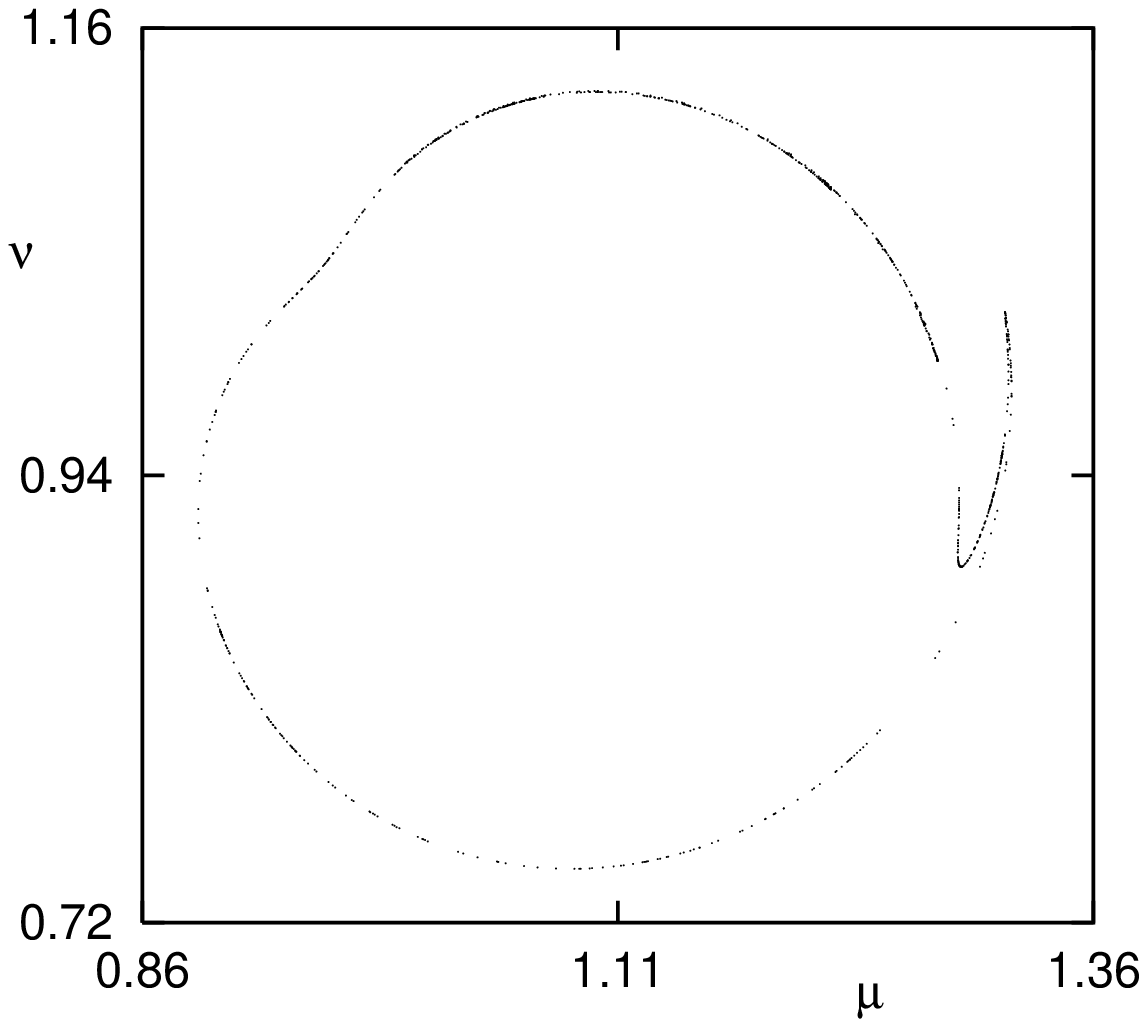,height=1.8in,width=1.6in}
}
\centerline{
\epsfig{file=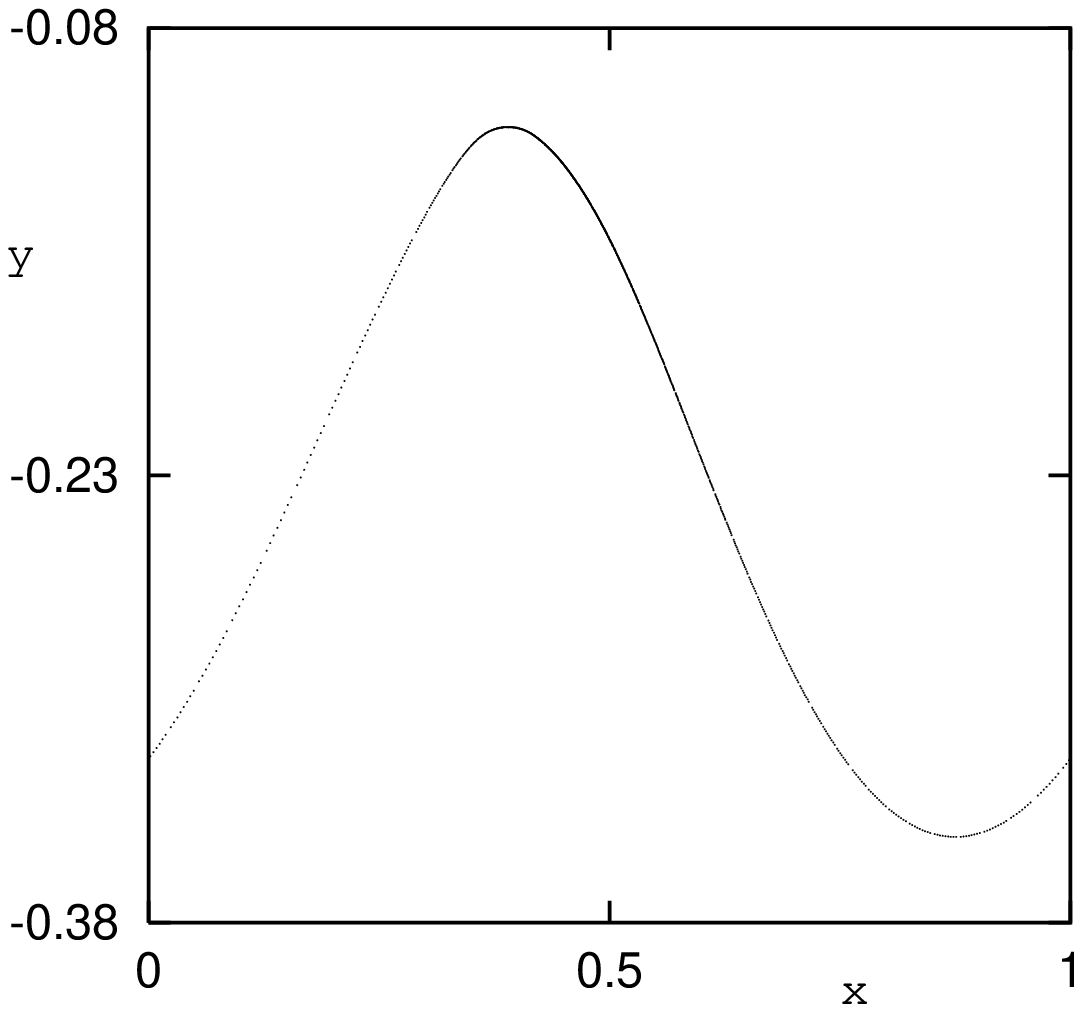,height=1.8in,width=1.6in}
\epsfig{file=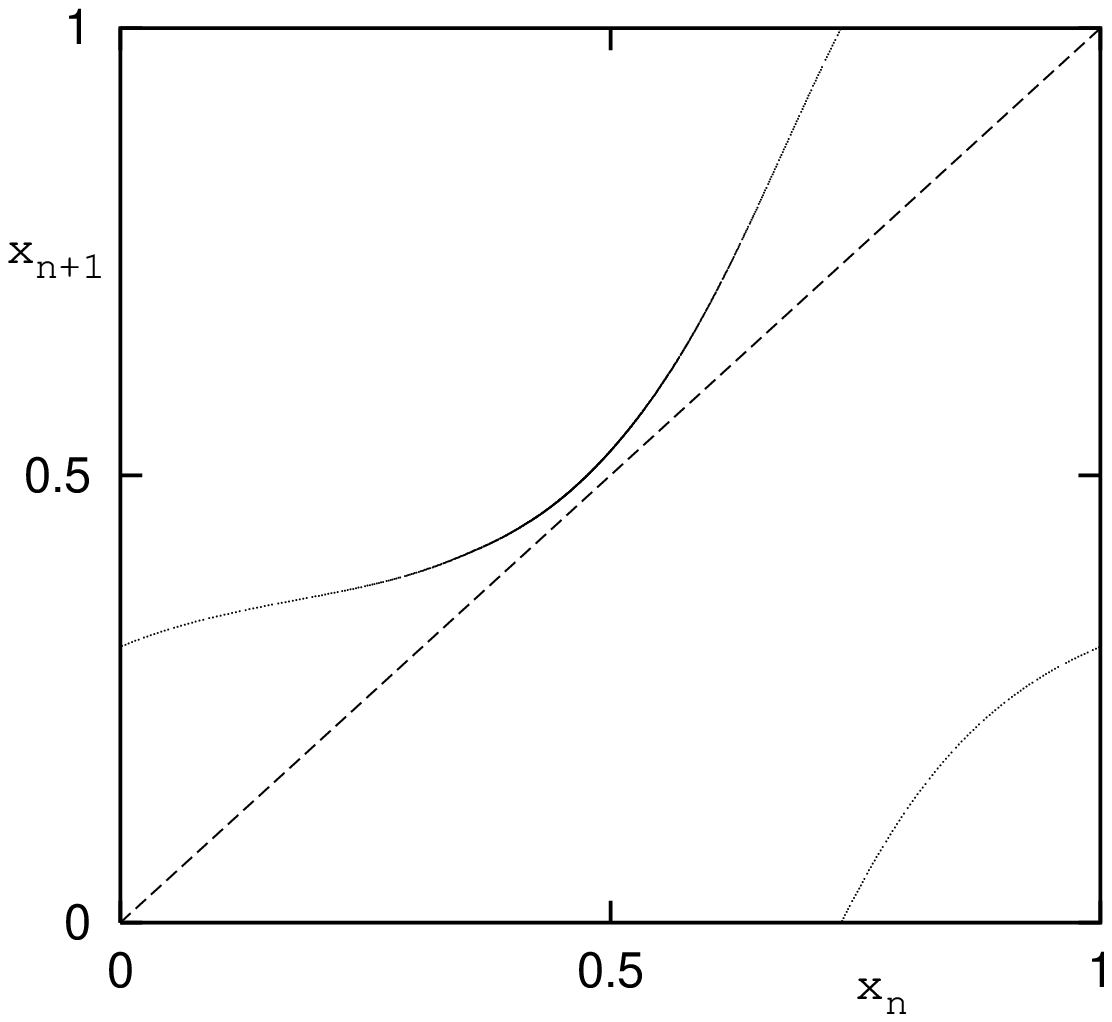,height=1.8in,width=1.6in}
\epsfig{file=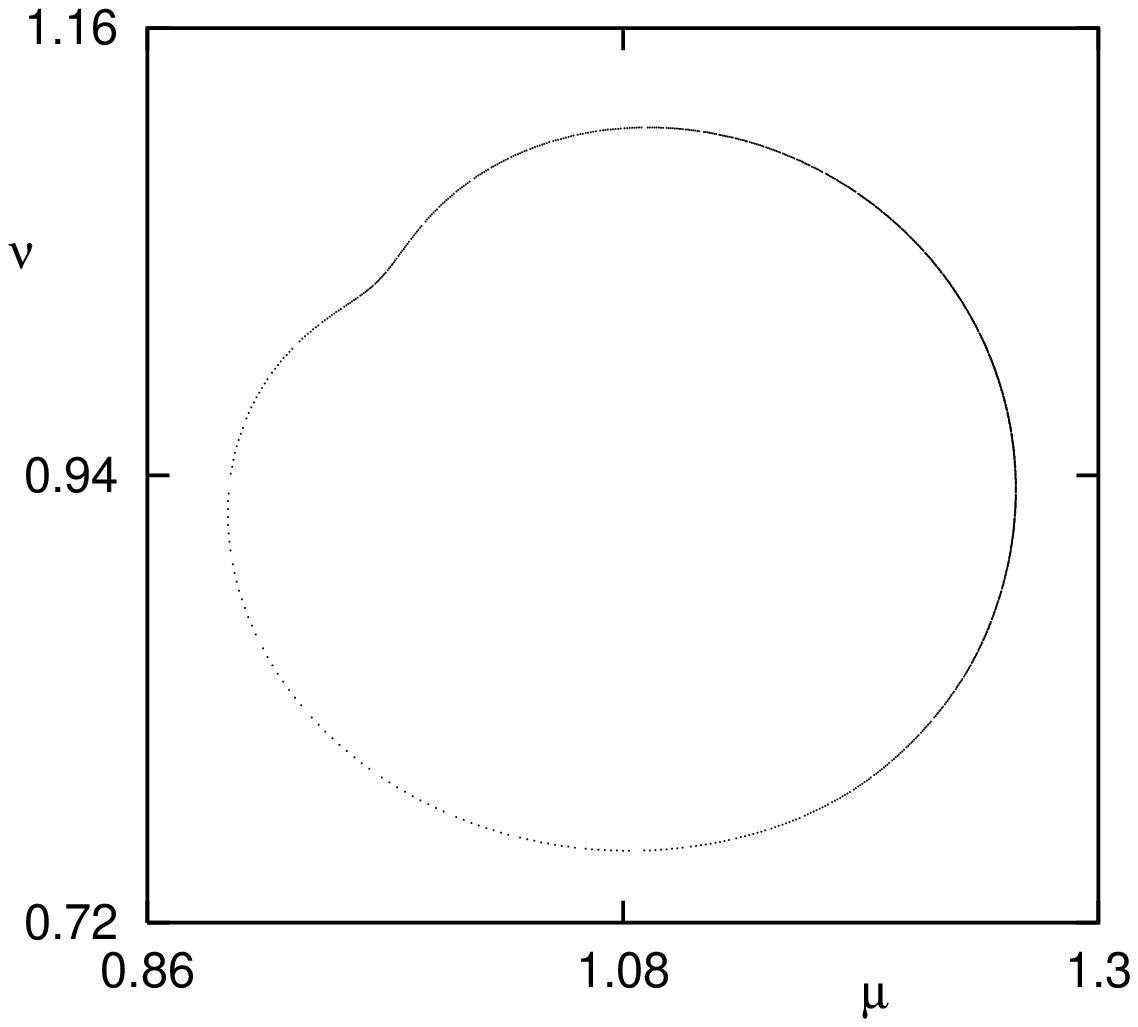,height=1.8in,width=1.6in}
}

\caption{Attractors shown in the $x - y,~ x_n - x_{n+1},$
and $\mu - \nu $ planes at $Q = 0.57, 0.76, 1.2577, 1.68,
{\rm and } ~ 2.0 $ from top to bottom for $r = 1.088, 
{\rm and } ~ a = 0.8 + 0.3Q.$}\label{jxlf1}
\end{figure}

To observe the dynamics changes for varying $Q$ and $a$ we juxtapose
the Poincar\'e sections (x, y) of attractors, 
the $x_n \rightarrow x_{n+1}$ first 
return map and the $\mu - \nu$ portraits where 
$\mu = 1 - (0.4 + y)\cos(2\pi x)$ and $\nu = 1 - (0.4 + y)\sin(2\pi x)$ at
$Q = 0.57, 0.76, 1.2577, 1.68, {\rm and } ~ 2.0 $ from top to bottom in
Fig.~\ref{jxlf1}.
Obviously, when $Q$ increases, 1) the portraits undergo the changes: 1D 
$\rightarrow$ 2D $\rightarrow$ 1D; 2) the return map $x_{n+1} - x_n$ shows:
a subcritical circle map without any decreasing branch $\rightarrow$
the appearance of decreasing branches  $\rightarrow$   
a subcritical circle map; 3) the dynamical
behavior exhibits: quasiperiodicity $\rightarrow$ chaos $\rightarrow$
quasiperiodicity.  
 
\begin{figure}
\centerline{\epsfig{file=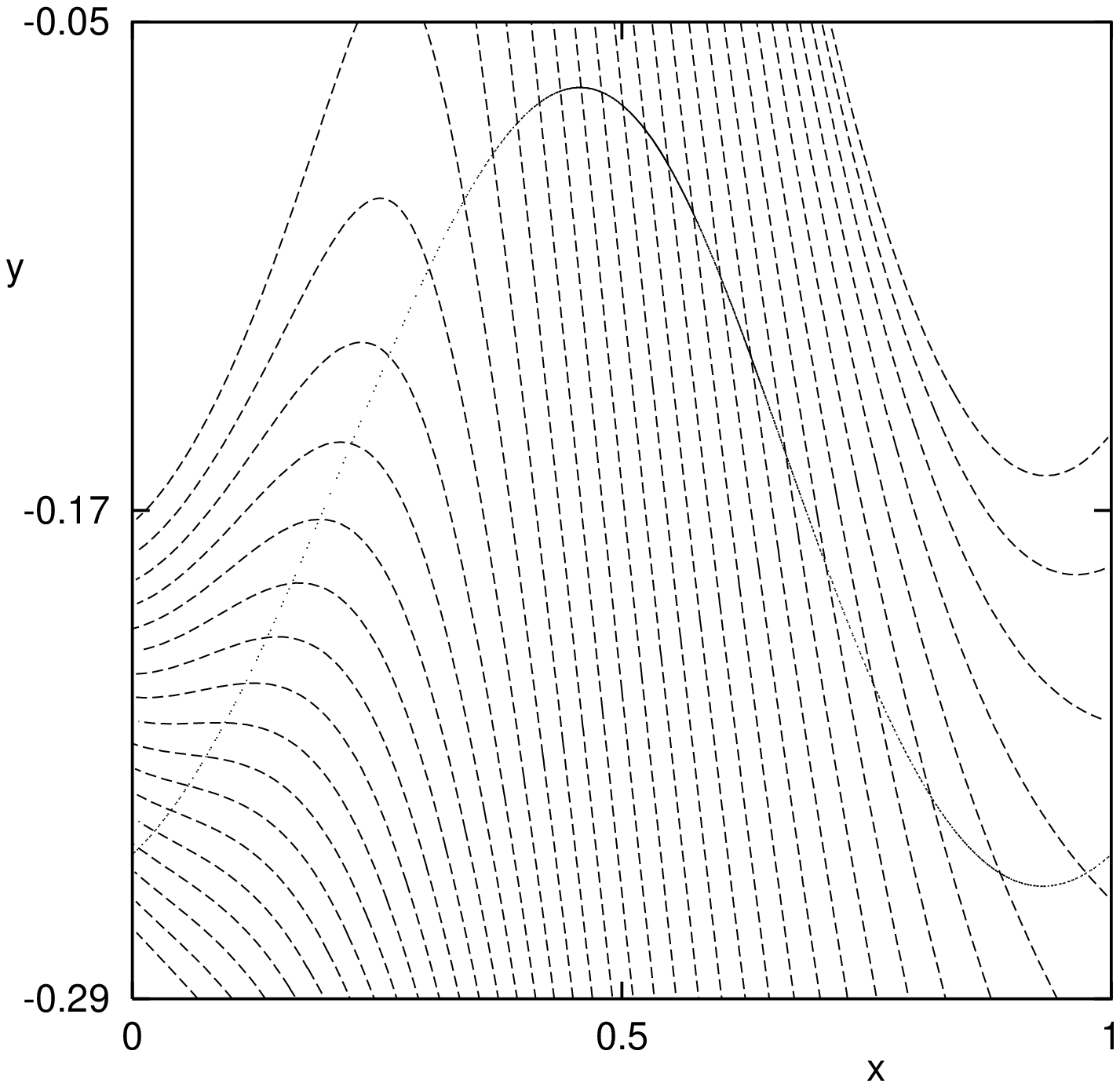,height=4.2in,width=4.0in}}
\caption{Attractor (dots) and forward foliations (dash curves)
for $Q = 0.57$.}\label{jxlf2}
\end{figure}

The quasiperiodic attractor and forward foliations at $Q = 0.57$
are shown in Fig.~\ref{jxlf2}, from which one can see clearly that there 
is no any tangency between the attractor (part of backward foliations)
and the forward foliations. As a matter of fact, when $Q$ increases
to a value the tangencies begin to appear, and correspondingly one can
obtain a
critical circle or annular map $x_{n+1} - x_n$. If $Q$ keeps increasing,
then
the map would be supercritical. Finally there are no tangency and the
map becomes subcritical again (see Fig.~\ref{jxlf1}). This shows the 
close connection between
the geometric properties (the tangencies) of an attractor and the dynamical
behavior (quasiperiodicity and chaos) of a system.     

\begin{figure}
\centerline{\epsfig{file=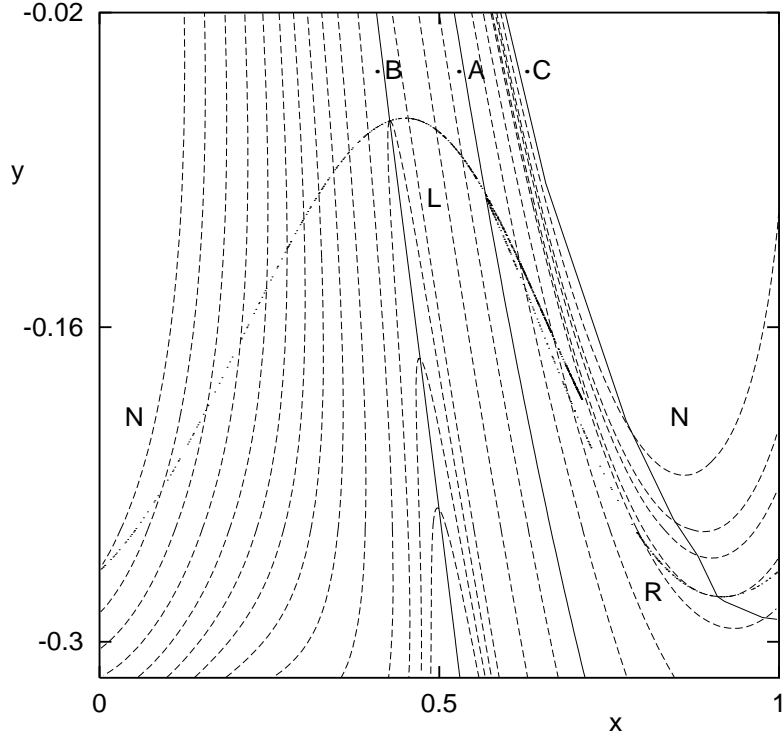,height=4.2in,width=4.0in}}
\label{f2}
\vspace {-.5cm}

\caption{The Poincar\'e map (dots) and forward foliations
(dash curves) at $Q = 0.76$. The primary partition lines 
$\bullet$B, $\bullet$C, and the pre-image $\bullet$A of 
$\bullet$B divide the attractor into three parts labeled
by the letters L, R and N.} \label{jxlf3}
\end{figure}

\begin{figure}
\centerline{\epsfig{file=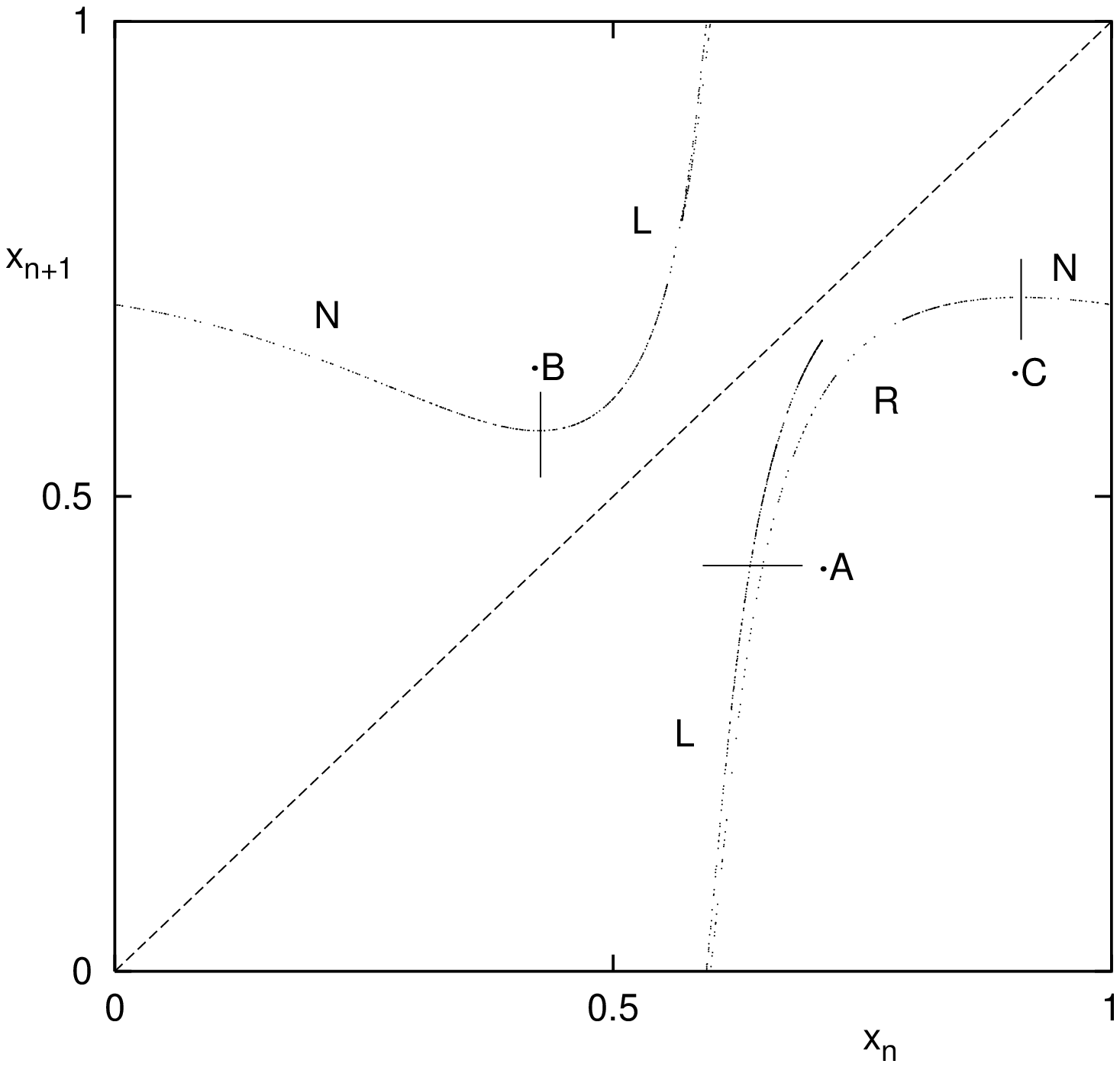,height=4.2in,width=4.0in}}
\caption{The $x_{n+1} - x_n $ first return map at $Q = 0.76$.}
\label{jxlf4}
\end{figure}

\begin{figure}
\centerline{\epsfig{file=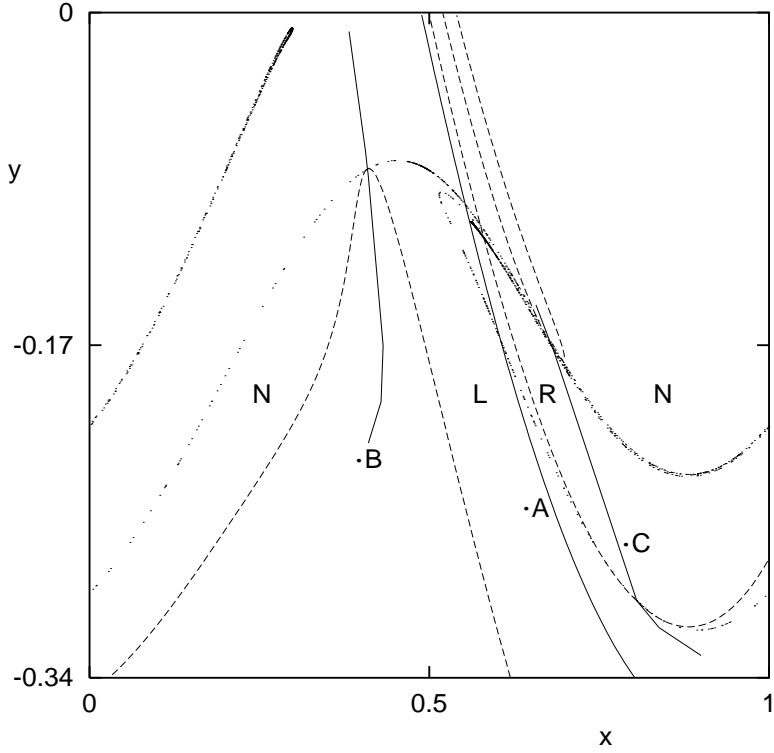,height=4.2in,width=4.0in}}
\caption{The chaotic attractor (dots), forward foliations (dashed lines),
and partition lines $\bullet$A, $\bullet$B, and $\bullet$C at 
$Q = 1.2577$.}\label{jxlf5}
\end{figure}

\begin{figure}
\centerline{\epsfig{file=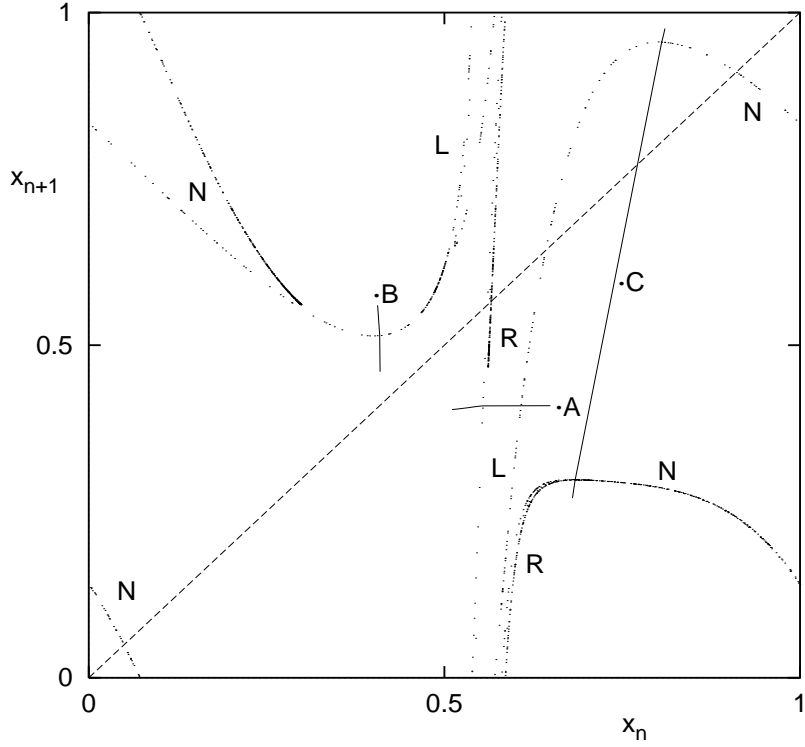,height=4.2in,width=4.0in}}
\caption{The $x_{n+1} - x_n $ first return map constructed from
Fig.~\ref{jxlf5}.}\label{jxlf6}
\end{figure}

\begin{figure}
\centerline{\epsfig{file=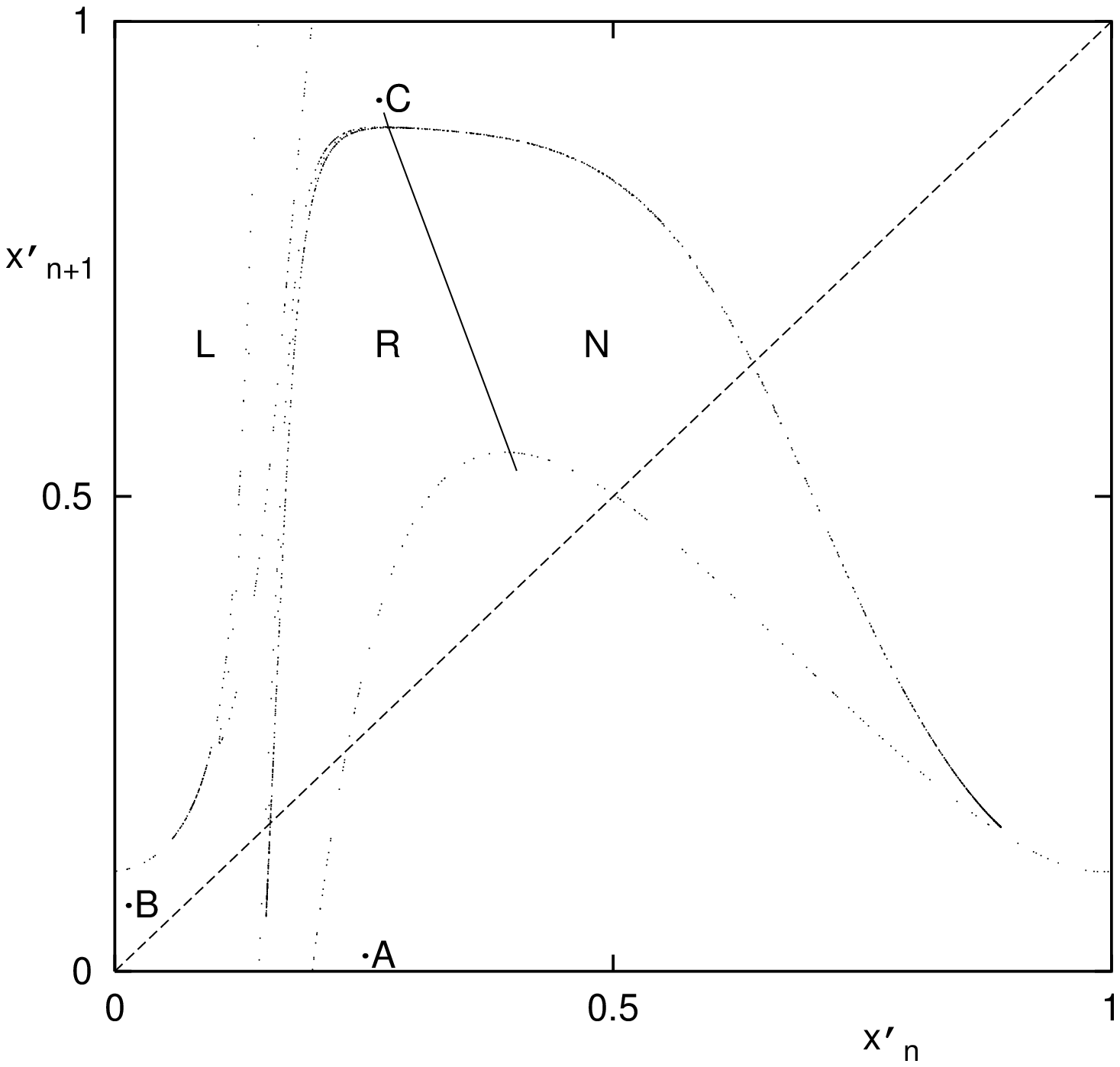,height=4.2in,width=4.0in}}
\caption{The $x'_{n+1} - x'_n $ swapped return map constructed
from Fig.~\ref{jxlf6}.}\label{jxlf7}
\end{figure}

In Fig.~\ref{jxlf3} we show the attractor and two primary
partition lines ($\bullet$B and $\bullet$C) on the background of the
forward foliations (dash curves) for $Q = 0.76$. The line marked with 
$\bullet$A is the pre-image of $\bullet$B. The areas in between these
lines are labeled by $\bullet R$, $\bullet L$, and $\bullet N$.
The corresponding first return map is plotted in Fig.~\ref{jxlf4}. 
The three monotone segments in Fig.~\ref{jxlf4} are
assigned the letters $L$, $R$, and $N$, in accordance with the
two-dimensional partitions in Fig.~\ref{jxlf3}.

A more interesting case is encountered at $Q = 1.2577$. The attractor,
three
forward foliations passing tangencies and the primary partition lines are 
shown in Fig.~\ref{jxlf5}, which manifestly exhibits 
two-dimensional feature. We also separately plot in Figs.~\ref{jxlf6}
and \ref{jxlf7} 
the $x_{n+1} - x_n $ first return map constructed
from Fig.~\ref{jxlf5} by using the $x$ coordinates and its swapped form 
$x'_{n+1} - x'_n $ with
\begin{equation}
x' = x + 1 - x_B \quad {\rm for }\quad x < x_B, \quad{\rm or}\quad 
x' = x - x_B \quad {\rm for }\quad x \geq x_B, 
\end{equation}
where $x_B$ is the $x$ coordinate of the attractor tangency on the
partition
line $\bullet B$ in Fig.~\ref{jxlf5}. 
Actually the partition lines $\bullet B$ and $\bullet A$ are not
respectively 
parallel to the $x_{n+1}$-axis and $x_n$-axis in 
Figs.~\ref{jxlf6} and \ref{jxlf7}.

\section{Ordering rule and admissibility condition}
\label{sec3}

After the partition lines are determined, each point, or its
orbit, may be encoded with
a doubly infinite symbolic sequence consisting of the letters $R$, $L$ and
$N$, say, $S = \cdots s_{\overline{m}} \cdots
s_{\overline{2}}s_{\overline{1}}\bullet
s_1 s_2 \cdots s_n \cdots $, where $s_n $ is the code for the $n$-th point
of the forward orbit, and $s_{\overline{m}}$ the code for the $m$-th point
of the
backward orbit. The ``present'' position is indicated by a solid dot, which
divides 
the doubly infinite sequence into two semi-infinite sequences, i.e., the
backward
sequence $\cdots s_{\overline{m}} \cdots
s_{\overline{2}}s_{\overline{1}}\bullet$
and the forward sequence $\bullet s_1 s_2 \cdots s_n \cdots $.

A metric representation for symbolic sequences can be introduced by 
assigning numbers in $[0,1]$ to forward and backward sequences. 
We first assign an integer $\epsilon_i = -1$ or $ 1$ to the symbol $s_i$
when it is
the letter $N$ or otherwise. Then we assign to the forward sequence 
$\bullet s_1 s_2 \cdots s_n \cdots $ the number
\begin{equation}
\label{eble3}
\alpha  = \sum_{i=1}^\infty \mu_i 3^{-i},  \ \ \ \ \
\end{equation}
where
\begin{equation}
\label{eble4}
\mu_i = \cases{0\cr 1\cr 2\cr} \quad {\rm for }\quad
s_i=  \cases{L\cr R\cr N\cr} \quad {\rm if} \quad
\prod_{j=1}^{i-1}\epsilon_j =1, 
\end{equation}
or,
\begin{equation}
\mu_i= \cases{2\cr 1\cr 0\cr} \quad {\rm for }\quad
s_i=  \cases{L\cr R\cr N\cr} \quad {\rm if} \quad
\prod_{j=1}^{i-1}\epsilon_j =-1. 
\end{equation}
Similarly, the $\beta $ assigned to the backward sequence 
$\cdots s_{\overline{m}} \cdots s_{\overline{2}}s_{\overline{1}}\bullet$
is defined by
\begin{equation}
\label{eble5}
\beta = \sum_{i=1}^\infty \nu_{\overline i} 3^{-i},\ \ \ \ \
\end{equation}
where
\begin{equation}
\label{eble6}
\nu_{\overline i} = \cases{0\cr 1\cr 2\cr} \quad {\rm for}
\quad s_i = \cases{R\cr N\cr L\cr}  \quad
 {\rm and }\quad \prod_{j=1}^{i-1}\epsilon_ {\overline j} = 1, 
\end{equation}
or
\begin{equation}
\nu_{\overline i} = \cases{2\cr 1\cr 0\cr} \quad {\rm for}
\quad s_i = \cases{R\cr N\cr L\cr}  \quad
 {\rm and }\quad \prod_{j=1}^{i-1}\epsilon_ {\overline j} = -1.
\end{equation}
According to the definition we have 
\begin{equation}
\label{eble7}
\begin{array}{l}
\alpha (\bullet NL^\infty) = \beta (L^\infty \bullet) =
1,\hspace{2.52cm} \alpha (\bullet L^\infty) = \beta (R ^\infty \bullet) =
0,\\
\alpha (\bullet NNL^\infty) = \alpha (\bullet RNL^\infty)=
2/3,\hspace{1.04cm}
\beta (R^\infty L\bullet) =\beta (R^\infty N\bullet) = 2/3, \\
\alpha (\bullet RL^\infty) = \alpha (\bullet LNL^\infty)=
1/3,\hspace{1.5cm}
\beta (L^\infty N\bullet) =\beta (L^\infty R\bullet) = 1/3.
\end{array}
\end{equation}
In this representation a bi--infinite symbolic sequence with the present dot
specified corresponds to a point in the unit square of the $\alpha - \beta$
plane, 
the so-called symbolic plane. In the plane, forward and backward foliations
become 
vertical and horizontal lines, respectively. We may define the ordering
rules of 
forward (or backward) sequences  according to their $\alpha$ (or $\beta$)
values. 
{}From Eqs. (\ref{eble3}-8) we then have
\begin{equation}
\label{eble8}
\begin{array}{lr}
\bullet EL\cdots <\bullet ER\cdots <\bullet EN\cdots,\hspace{1cm}
\bullet OL\cdots >\bullet OR\cdots >\bullet ON\cdots,\\
\end{array}
\end {equation}
and
\begin{equation}
\label{ess18}
\begin{array}{lr}
\cdots RE\bullet <\cdots NE\bullet <\cdots LE\bullet,\hspace{1cm}
\cdots RO\bullet >\cdots NO\bullet >\cdots LO\bullet,\\
\end{array}
\end{equation} 

where the finite strings $E$ and $O$ consist of letters $L$, $R$ and $N$
and contain an 
even and odd number of letter $N$, respectively. This ordering rule is
similar to
that for sequences of the dissipative standard map at some values of the 
parameters \cite{lzh96}.

\begin{figure}
\centerline{\epsfig{file=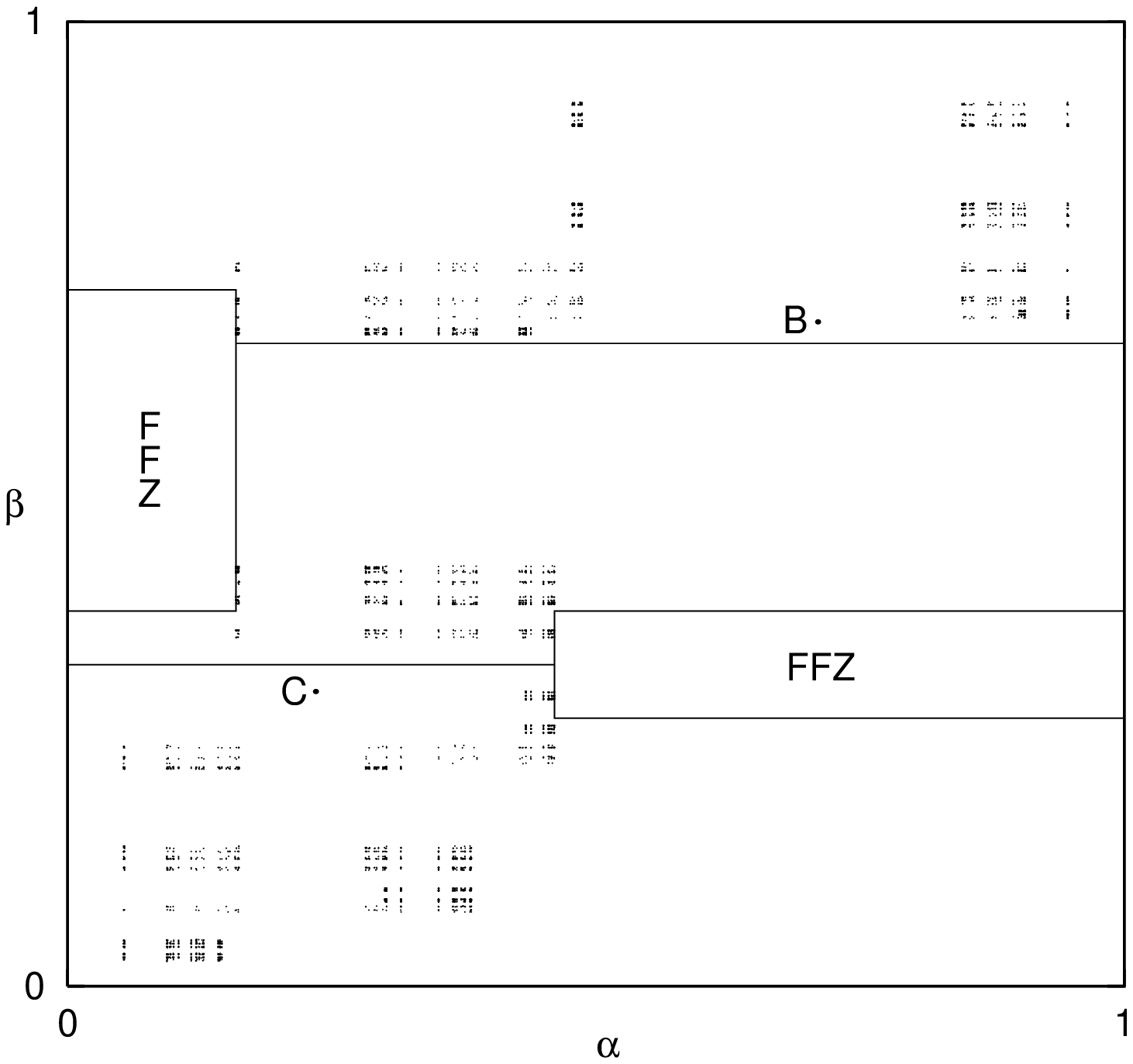,height=4.2in,width=4.0in}}
\caption{The symbolic plane at $Q = 0.76$. 10000 points of real
orbits generated from the Poincar\'e map are also shown together with the
FFZ in which no point falls.}\label{jxlf8}
\end{figure}

\begin{figure}
\centerline{\epsfig{file=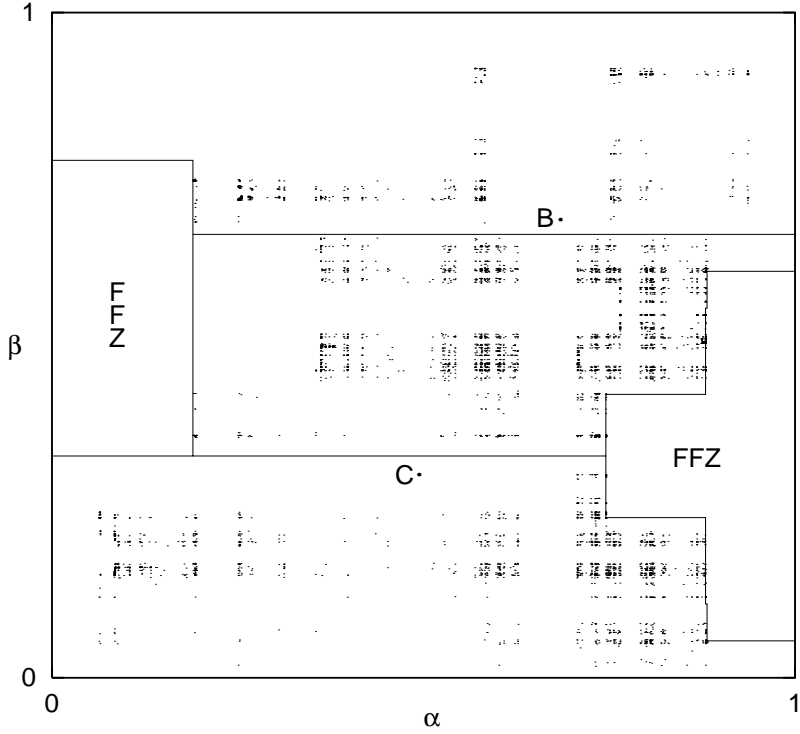,height=4.2in,width=4.0in}}
\caption{The symbolic plane at $Q = 1.2577$. Together with the FFZ,
10000 points representing real orbits are drawn. None of them
falls inside the FFZ.}\label{jxlf9}
\end{figure}

\begin{figure}
\centerline{\epsfig{file=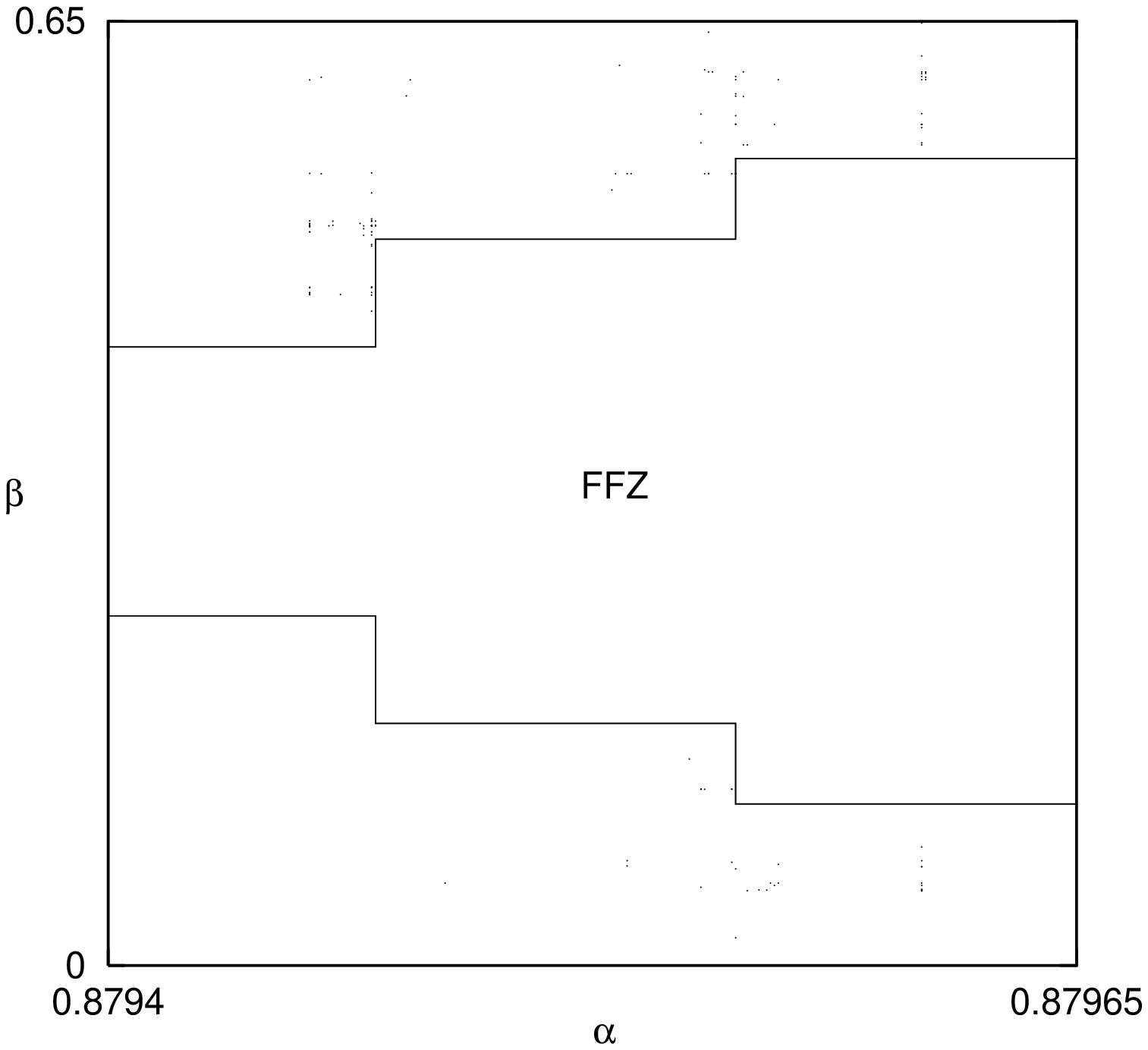,height=4.2in,width=4.0in}}
\caption{A blow-up of the symbolic plane Fig.~\ref{jxlf9} in the intervals
$\alpha = [0.8794, 0.87965]$ and $\beta = [0, 0.65]$.}\label{jxlf10}
\end{figure}

When foliations are well ordered, the geometry of a tangency places a
restriction on allowed
symbolic sequences. A point on the partition line $C\bullet $ (image of
$\bullet C$) may 
symbolically be represented as $QC\bullet P$. The rectangle enclosed by the
lines 
$QN\bullet $, $QR\bullet $, $\bullet P$, 
and $\bullet NL^\infty $ forms a forbidden zone (FZ) in the symbolic plane.
Therefore, a
symbolic sequence $IJ$ with $I\bullet $ between $QN\bullet $ and $QR\bullet
$,
and at the same time $\bullet J>\bullet P$ must  be forbidden by the
tangency $QC\bullet P$. 
In the symbolic plane the sequence $IJ$ corresponds to a point inside the
forbidden zone of
$QC\bullet P$. Similarly, $UB\bullet V$ stands for a tangency on the
partition line
$B\bullet $ (image of $\bullet B$). The lines $UL\bullet$, $UN\bullet$,
$\bullet V$ and
$\bullet L^\infty $ enclose a rectangle FZ in the symbolic plane. Any
sequence $KT$ with
$K\bullet $ between $UL\bullet $ and $UN\bullet $ while $\bullet T <
\bullet V$ is forbidden
by the $UB\bullet V$. Each tangency point on a partition line rules out a
rectangle in 
the symbolic plane. The union of the forbidden rectangles, determined from
one and the 
same partition line, forms the fundamental forbidden zone (FFZ),
a boundary of which is the so-called pruning front. 
Consider a finite set of tangencies
\{$Q_iC\bullet P_i$\} (or \{$U_jB\bullet V_j$\}). If the shift of a
sequence 
$\cdots s_{k-2}s_{k-1} \bullet s_k s_{k+1}\cdots $ 
satisfies the condition that the backward sequence $\cdots s_{k-2}s_{k-1}
\bullet $ is not
between $Q_i N\bullet $ and $ Q_i R\bullet $ (or $U_j N\bullet $ and $ U_j
L\bullet $)
, and at the same time $\bullet P_i>\bullet s_k s_{k+1}\cdots $
(or $\bullet V_j<\bullet s_k s_{k+1}\cdots $) 
for some $i$ ($j$), then this shift is not forbidden by any tangencies of
$C\bullet$ or 
$B\bullet$, owing to  the property
of well-ordering of foliations. Thus, we may say that the shift is allowed
according to
that tangency. A necessary and sufficient condition for a sequence to be
allowed is that all
of its shifts are allowed according to the two sets of tangencies.
To check the admissibility condition, we consider again the two cases $Q =
0.76$ and  
$Q = 1.2577$, and draw $10000$
points representing real sequences generated from the Poincar\'e map
together with
the FFZ in the symbolic plane, as seen in 
Figs.~\ref{jxlf8} and \ref{jxlf9}. One can see that
the FFZ indeed 
contains no point of allowed sequences. A blow-up of 
the right-hand side pruning front in Fig.~\ref{jxlf9} is displayed in 
Fig.~\ref{jxlf10}. The
structure 
means two-dimensional feature, related to the two tangent points in the
upper part 
of attractor on the partition line $\bullet C$ (see Fig.~\ref{jxlf5}). 
We shall use the two
tangencies to make 2D analysis of periodic sequences later on (see T$_3$
and T$_4$ in the next section).

\section{Unstable periodic orbit sequences}
\label{sec4}

 The attractor at $Q = 0.76$ does
not show much two--dimensional nature, so the reduction to symbolic
dynamics of
one-dimensional circle map may capture much of the essentials. We start from
this simple case. The attractor
resembles that of a one--dimensional circle map except for a segment with
two
sheets, one of which without the $N$ part (
see Figs.~\ref{jxlf3} and \ref{jxlf4}). 
{}From the two primary partition lines $\bullet C$ and $\bullet B$ in 
Fig.~\ref{jxlf3}, 
we get the following sequences for attractor points:

$$
\begin{array}{rc}
{\rm T_1:}& \cdots NN\cdots NNAB\bullet LRRLNLNLNLRRRLLRRR \cdots ,\\
{\rm T_2:}& \cdots NN\cdots NNLC\bullet RRLRLNLRRRLRLNRLNL \cdots . \\ 
\end{array}
$$

In order to reduce the two-dimensional attractor to one-dimensional return
map, 
we need to determine two kneading sequences $K_B$ and $K_C$. They are the
forward
sequences of ${\rm T_1}$ and ${\rm T_2}$, respectively. 
\begin{equation}
\label{ess21}
\begin{array}{rcl}
{\rm K_B} &=& LRRLNLNLNLRRRLLRRR \cdots ,\\
{\rm K_C} &=& RRLRLNLRRRLRLNRLNL \cdots .\\
\end{array}
\end{equation}
Compared with the original 2D map, the 1D circle map given by these 
${\rm K_B}$
and ${\rm K_C}$ puts less constraints on allowed orbits. Since the
attractor
has only one sheet crossing each primary 
partition line nearly no difference between the 1D and 2D maps can be 
recognized if the sequences of short periodic orbits are concerned.

The knowledge of the two kneading sequences (\ref{ess21}) determines
everything
in the symbolic dynamics of the circle map \cite{z9194}. For example,
one may define a {\it rotation number} $W$, also called a {\it winding
number},
for a symbolic sequence by counting the weight of letters $R$ and $N$ in
the 
total number $n$ of all letters:
\begin{equation}
\label{ess22}
W=\lim_{n\rightarrow\infty}{\frac 1 n}{\rm
(Number\,\,\,of\,\,\,R\,\,\,and\,\,\,N)}.
\end{equation}
Chaotic regime is associated with the existence of a rotation interval, a
closed interval in the parameter plane \cite{ito81}. Within a rotation
interval there must be well-ordered orbits. We can construct some of these
well-ordered sequences explicitly, knowing the kneading sequences 
${\rm K_B}$ and ${\rm K_C}$.

 In our case it can be verified that the ordered
periodic orbits $(RL)^\infty$ and $(RRL)^\infty$ are admissible.
These two sequences have rotation numbers 1/2 and 2/3, so the rotation
interval of the circle map contains [1/2, 2/3], inside which there are
rational rotation numbers 3/5, 4/7 and 5/8 with denominators up to 8.
Their corresponding ordered orbits are $(R^2LRL)^\infty$,
$[R^2L(RL)^2]^\infty$ and $[(R^2L)^2RL]^\infty$.
A very easy way to construct a longer well-ordered periodic sequence with a
given 
rational rotation number from two shorter well-ordered periodic sequences
can be
found in Fig.~\ref{jxlf4}.9 of Ref.\cite{h89}. Take, for example $W = 3/5$,
$${\frac 3 5}={\frac 2 3} \oplus {\frac 1 2} = {\frac {2 + 1} {3 + 2}},$$
$$RRLRL = RRL + RL.$$

We can further construct not-well-ordered sequences from well-ordered ones
by
the following transformation. One notes that in Fig.~\ref{jxlf4} the 
lower limit of 
$\bullet A$ is
the greatest point on the subinterval $L$, while the upper limit of 
$\bullet A$ is the
smallest $R$. When $\bullet A$ is crossed by a continuous change of initial
points the corresponding symbolic sequences must change as follows:
$${\rm greatest}\; LN\cdots\rightleftharpoons {\rm smallest}\;RL\cdots.$$
Similarly, on crossing $\bullet C$ another change of symbols takes
place:
$${\rm greatest}\; R \rightleftharpoons {\rm smallest}\; N. $$
Neither change has any effect on rotation numbers. As an example, starting 
with the ordered period~7 orbit $[R^2(LR)^2L]^\infty$ we obtain
$$ RRLRLRL \rightarrow NRLRLRL \rightarrow NRLRLNL \rightarrow NRLNLNL
\rightarrow NRLNRLL $$
$$\rightarrow NRRLRLL \rightarrow RRRLRLL \rightarrow
RRRLLNL \rightarrow NRRLLNL \rightarrow NNRLLNL $$ and
$$ NRLRLNL \rightarrow NRLRRLL \rightarrow RRLRRLL $$
as candidates for the fundamental strings in not-well-ordered sequences of
period~7. Among these sequences, $(NNRLLNL)^\infty $ and 
$(RRLRRLL)^\infty $ are forbidden by
$K_C$ and $K_B$, respectively.

In this way we have determined all periodic sequences up to period 8,
allowed
by the two kneading sequences (\ref{ess21}). The result 
is summarized in Table \ref{jxlt1}. We have examined the admissibility of all
these sequences by checking if their all shifts fall into the FFZ in the
symbolic plane of Fig.~\ref{jxlf8}. They turn out to be totally allowed. 
In fact, 
by determining the symbolic sequence of every point in the attractor, 
we have numerically found all these orbits easily and listed the coordinate
of the first letter in a sequence in Table \ref{jxlt1}.

\begin{table}[H]
\begin{center}
\caption{Allowed unstable periods up to 8 for $Q=0.76$. The
$(x, y)$ is the coordinate of the first letter in a sequence.
Only non--repeating strings of the sequences are given. $P$ 
denotes the period and $W$ the rotation number.}\label{jxlt1}
\          \\
\begin{tabular}{cclcc}
\hline
$P$ & $W$& Sequence & $x$ & $y$ \\
\hline
2  & 1/2  & $ RL$ & 0.680655071 & -0.184044078 \\

2  & 1/2  & $ NL$ & 0.269335571 & -0.128679398  \\ 

3  & 2/3  & $ RLR $ & 0.666418156 & -0.163903964 \\

3  & 2/3  & $ RLN $ & 0.679742958 & -0.172914732  \\

4  & 2/4  & $ NRLL $ & 0.181303156 &  -0.181685939 \\ 

5  & 3/5  & $ RRLRL $ & 0.754061730 & -0.232614919 \\ 

5  & 3/5  & $ NRLRL $ & 0.153005251 & -0.198571214 \\

5  & 3/5  & $ NRLNL $ & 0.069305032 & -0.242724621 \\

5  & 3/5  & $ NRRLL $ & 0.004459397 & -0.266933087 \\

5  & 3/5  & $ RRRLL $ & 0.818461882 & -0.262557048 \\

5  & 3/5  & $ RRLNL $ & 0.805821652 & -0.257789205 \\

6  & 3/6  & $ NRLRLL $ & 0.162899971 & -0.192734200 \\

6  & 3/6  & $ NRLLNL $ & 0.176235170 & -0.184748719 \\

6  & 4/6  & $ RRRLRL $ & 0.876708867 & -0.277086130 \\

6  & 4/6  & $ NRRLRL $ & 0.941828668 & -0.278808659 \\

6  & 4/6  & $ RRLNRL $ & 0.781248148 & -0.246935351 \\

7  & 4/7  & $ RRLRLRL $ & 0.753419555 & -0.232248988 \\

7  & 4/7  & $ NRLRLRL $ & 0.159195421 & -0.194929335 \\

7  & 4/7  & $ RRLLNRL $ & 0.744687447 & -0.227153254 \\

7  & 4/7  & $ NRLRLNL $ & 0.151174731 & -0.199641399 \\ 

7  & 4/7  & $ NRLLNRL $ & 0.185646405 & -0.179051255 \\

7  & 4/7  & $ NLNLNRL $ & 0.259007234 & -0.134664855 \\

7  & 4/7  & $ RRRLLNL $ & 0.850218105 & -0.272009133 \\

7  & 4/7  & $ NRRLLNL $ & 0.981762965 & -0.272665749 \\

7  & 4/7  & $ RRRLRLL $ & 0.858495739 & -0.273871030 \\

7  & 4/7  & $ NRRLRLL $ & 0.964283072 & -0.275994309 \\

8  & 4/8  & $ NRLRLRLL $ & 0.160408189 & -0.194212064 \\

8  & 4/8  & $ NRLRLLNL $ & 0.162092260 & -0.193213633 \\

8  & 4/8  & $ NRLLNLNL $ & 0.177466432 & -0.184005818 \\

8  & 5/8  & $ RRLRRLRL $ & 0.759627678 & -0.235734676 \\

8  & 5/8  & $ NRLRRLRL $ & 0.143963347 & -0.203824391 \\

8  & 5/8  & $ NRLNRLRL $ & 0.093629167 & -0.231107382 \\

8  & 5/8  & $ NRRLRLRL $ & 0.953347043 & -0.277574465 \\

8  & 5/8  & $ RRRLRLRL $ & 0.869038801 & -0.275879784 \\

8  & 5/8  & $ NRLNLNRL $ & 0.070703636 & -0.242089800 \\

8  & 5/8  & $ NRLNRRLL $ & 0.111829461 & -0.221687365 \\

8  & 5/8  & $ NRLRRRLL $ & 0.131212194 & -0.211077481 \\

8  & 5/8  & $ RRLNLNRL $ & 0.792783112 & -0.252286399 \\

8  & 5/8  & $ RRLNRLRL $ & 0.782318121 & -0.247450423 \\

8  & 5/8  & $ NLNRRLRL $ & 0.351020175 & -0.087984499 \\

8  & 5/8  & $ NLRRRLRL $ & 0.353562509 & -0.086982096 \\

\hline
\end{tabular}
\end{center}
\end{table}

For the  more interesting case $Q = 1.2577$, based on Fig.~\ref{jxlf5} 
we list the following five tangencies along the
$B\bullet$ and $C\bullet$ lines:
$$
\begin{array}{l}
{\rm T_1:} \cdots L\cdots LLAB\bullet LRNNRLLNNNNNRLRLLN \cdots,\\
{\rm T_2:} \cdots N\cdots NNRC\bullet NRLLRNNNNRLRLRNNRN \cdots,\\
{\rm T_3:} \cdots L\cdots LLNC\bullet NRLLNNNRNRNNRNRNNR \cdots,\\
{\rm T_4:} \cdots N\cdots NNNC\bullet NRLLNNNRLRNNNNRLLN \cdots,\\
{\rm T_5:} \cdots N\cdots NRLC\bullet NNNNRLRLNRNRLNNNRN \cdots.
\end{array} 
$$
{}From T$_1$ and T$_2$ whose forward sequence is the greatest
among the tangencies along $C\bullet$, we get 
$$
\begin{array}{lcc}
{\rm K_B} &=& LRNNRLLNNNNNRLRLLN \cdots,\\
{\rm K_C} &=& NRLLRNNNNRLRLRNNRN \cdots.\\
\end{array}
$$

\begin{table}[H]
\begin{center}
\caption{Allowed unstable periods up to 7 for $Q=1.2577$ in the 1D case;
those with an asterisk are forbidden by 2D tangency T$_5$.}\label{jxlt2}
\         \\
\begin{tabular}{ccl}
\hline
$P$ & $W$& Sequences \\
\hline
1  &  1/1 & $   R ~~ N$ \\
2  &  2/2 & $  RN ~~  $    \\
2  & 1/2  & $RL ~~ NL$\\
3  &  3/3 & $ NRR~~  RNN $ \\
3  & 2/3  & $ RLR~~ RLN $\\
4  &  4/4 &$ RRRN ~~ RRNN~~  RNNN $ \\ 
4 &  2/4  &$ NRLL  $   \\
4  &  3/4 &$ RRRL~~  NRRL~~ NNRL^\ast $ \\
5  &  5/5 &$ RRRRN ~~ RRRNN~~  RRNNN~~  RNNNN ~~ RNNRN~~ RRNRN $\\
5 &  3/5  & $ RRLRL~~ NRLRL~~ NRLNL~~ NRRLL~~ NNRLL^\ast $ \\
5 &   4/5 & $ RRRRL~~ NRRRL~~ NNRRL~~ NNNRL~~ RNRRL~~ RNRLN~~ RNRLR$\\  
6 &   6/6 & $ RRRRRN ~~ RRRRNN~~ RRRNNN~~ RRNNNN~~ RNNNNN~~RRRNRN $\\
6 &   6/6 & $ RRNNRN~~ RNNNRN~~ NNRRNR $\\  
6 &  3/6 & $ RLRLLN ~~RLLNLN $ \\ 
6  & 4/6 & $ RRRLRL~~ RLNRRL~~ RLNNRL^\ast~~ NLNNRL^\ast~~ RLRRLN~~ RRLNLN
$\\ 
6  &  5/6 & $ RRRRRL~~ NRRRRL~~ NNRRRL~~ NNNRRL~~ NNNNRL^\ast ~~RNRRRL $\\
6  &  5/6 & $ NRNRRL~~ RRNRRL~~ NNRNRL^\ast~~ RNRNRL^\ast~~RRRNRL~~ RRNNRL
$\\ 
6  &  5/6 & $ RNNNRL^\ast $\\
7  &  7/7 & $ RRRRRRN~~ RRRRRNN~~ RRRRNNN~~ RRRNNNN~~ RRNNNNN $\\ 
7  &  7/7 & $ RNNNNNN~~ RRRRNRN~~ RRRNNRN~~ RRNNNRN~~ RRNRNRN $\\
7  &  7/7 & $ RNNNNRN~~ RNNRNRN~~ RRNRNNN~~ RNNRRRN $\\  
7 &  4/7  & $ RRLRLRL~~ NRLRLRL~~ RRLLNRL~~ NRLRLNL~~ NRLLNRL~~ NLNLNRL $
\\
7  &  4/7 & $ RRLLNLN~~ RRLRLLN~~ NRLLNLN^\ast $\\
7  & 5/7  & $ RRLRRLR ~~RRLRRLN~~ RRLNRLN ~~NRLNRLN~~ RRLNLNN^\ast~~
NRLNLNN $\\ 
7  &  6/7 & $ RRRRRRL~~ NRRRRRL~~ NNRRRRL~~ NNNRRRL~~ NNNNRRL~~ NNNNNRL $\\
7  &  6/7 & $ RNRRRRL~~ NRNRRRL~~ RRNRRRL~~ NNRNRRL~~ RNRNRRL~~ RRRNRRL $\\
7  &  6/7 & $ NNNRNRL~~ NRNRNRL~~ NRRRNRL~~ RRRRNRL~~ NRRNRRL~~ RNNNNRL $\\
7  &  6/7 & $ RNNRNRL~~ RRNRNRL$\\
\hline
\end{tabular}
\end{center}
\end{table}

For the 1D circle map, we have determined all allowed periodic sequences up
to period~7, which are listed in Table \ref{jxlt2}. 
We have examined their admissibility
by using the tangencies of the 2D Poincar\'e map and found that ten of 
these cycles are now forbidden by the tangency T$_5$. An asterisk denotes
those forbidden sequences in Table \ref{jxlt2}. 
The allowed periodic orbits have 
been located numerically.

\section{Period windows}
\label{sec5}

\begin{figure}
\centerline{\epsfig{file=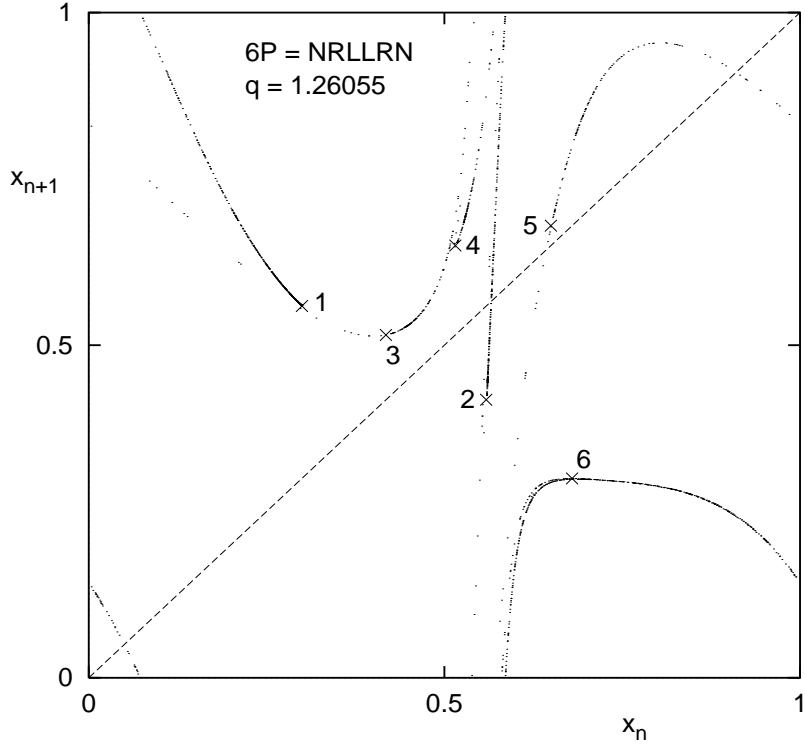,height=4.2in,width=4.0in}}
\caption{Stable period 6 NRLLRN at $Q = 1.26055$ and the chaotic
attractor at $Q = 1.26005$ in the $x_{n+1} - x_n$ map.}\label{jxlf11}
\end{figure}

 So far we have discussed only unstable periodic orbits which
are embedded in
a chaotic attractor for fixed parameters. In fact, the symbolic dynamics we
have constructed is also capable of treating stable periodic orbits
appearing
while parameters vary once the partition lines are determined in the
Poincar\'e map. But a more convenient way to determine the symbols of a
periodic
orbit is to use the $x_{n+1} - x_n$ first return map. Take, for instance a
stable
period 6 at $Q = 1.26055$. We show it and the chaotic
attractor at $Q = 1.26005$ in the $x_{n+1} - x_n$ map in Fig.~\ref{jxlf11}.
Comparing Fig.~\ref{jxlf11} with Fig.~\ref{jxlf6}, one can easily 
find 1=$N$, 4=$L$, and 5=$R$ 
for the two chaotic attractors look very similar. To tell the symbols of 
the rest points we need further to specifically locate the tangencies close
to 3 and 6 on partition lines $\bullet B$ and $\bullet C$. As a result,
we found a tangency (0.409421208764, 0.513670945215) on the left side of 
point 3 (0.417925742243, 0.515459734623) and the other 
(0.678580462575, 0.299669053868) also on the left side of  
6 (0.679571155041, 0.299669061327). Hence, the letters for 
points 2, 3, and 6 
should be $R$, $L$, and $N$, respectively. We have determined all stable 
periodic words with period up to 7 encountered as $Q$ is varied from
0.78058
to 1.32280 with an increment 0.00001, as listed in Table \ref{jxlt3}. 
Naturally, those 
periods with window width less than 0.00001 would be missing.        
In Table \ref{jxlt3} 
we also list the changes of period words with $Q$, which show
the ordering properties:
The words in the form $\cdots C$ would undergo the change from smaller to
bigger as $Q$
increases where $C$ corresponds to the tangency in the upper branch of
attractor
on the partition line $\bullet C$ (see Fig.~\ref{jxlf5}); while the 
words in forms 
$\cdots C^\dagger$ and $\cdots AB$ would do the opposite, i.e., change from
bigger to
smaller, where $C^\dagger$ is the tangency in the lower branch of attractor
along the $\bullet C$ (see Fig.~\ref{jxlf5}).

\begin{table}[H]
\begin{center}
\caption{Period windows with period up to 7
alone the line $a=0.8 + 0.3Q$.}\label{jxlt3}
\           \\
\begin{tabular}{cll}
\hline
$P$ & Range in $Q$ & Word and its change with $Q$\\
\hline
1 & 0.78058 - 1.03032 & $ C~~(R\rightarrow C\rightarrow N) $\\ 
2 & 1.03033 - 1.18422 & $ NC~~(NN\rightarrow NC\rightarrow NR) $\\
4 & 1.18423 - 1.20583 & $ NRNC~~(NRNR\rightarrow NRNC\rightarrow NRNN) $\\ 
6 & 1.21727 - 1.21783 & $ NRNNNC~~(NRNNNR\rightarrow NRNNNC\rightarrow
NRNNNN)$\\
7 & 1.22590 - 1.22596 & $ NRNNNNC~~(NRNNNNN\rightarrow NRNNNNC\rightarrow
NRNNNNR) $\\
5 & 1.23171 - 1.23228 & $ NRNNC~~(NRNNN \rightarrow NRNNC \rightarrow
NRNNR) $\\
7 & 1.23667 - 1.23673 & $ NRNNRNC~~(NRNNRNR \rightarrow NRNNRNC \rightarrow
NRNNRNN) $\\
3 & 1.24492 - 1.24591 & $ NRC~~(NRN\rightarrow NRC \rightarrow NRR) $\\
6 & 1.24592 - 1.24631 & $ NRRNRC~~(NRRNRR \rightarrow  NRRNRC \rightarrow
NRRNRN) $\\
5 & 1.250853 - 1.25087 & $ NRRNC~~(NRRNR \rightarrow NRRNC \rightarrow
NRRNN) $\\
7 & 1.255025 - 1.25503 & $ NRLNNNC~~(NRLNNNR \rightarrow NRLNNNC
\rightarrow NRLNNNN) $\\
7 & 1.25645 - 1.256457 & $ NRLRNNC~~(NRLRNNN \rightarrow NRLRNNC
\rightarrow NRLRNNR) $\\
6 & 1.26055 - 1.26173& $ NABLRC~~(NRLLRN\rightarrow NRLLRC\rightarrow
NRLLRR\rightarrow$\\
  &                   & $
 NABLRR\rightarrow NLNLRR\rightarrow NLNLRC \rightarrow NLNLRN)$\\ 
6 & 1.26735 - 1.26738 & $ LRRNAB~~(LRRNLN \rightarrow LRRNAB\rightarrow
LRRNRL) $\\
4 & 1.27539 - 1.27692 & $ NLNC~~(NLNR \rightarrow NLNC \rightarrow NLNN)
$\\
7 & 1.28907 - 1.28909 & $ NLNNNRC~~(NLNNNRR \rightarrow NLNNNRC \rightarrow
NLNNNRN) $\\
7 & 1.31289 - 1.31293 & $ NLRNNRC~~(NLRNNRN \rightarrow NLRNNRC \rightarrow
NLRNNRR) $\\
7 & 1.31700 - 1.31713 & $ NNRNNLC^\dagger~~(NNRNNLN \rightarrow
NNRNNLC^\dagger 
\rightarrow NNRNNLR) $\\
3 & 1.32043 - 1.32199 & $ LAB~~(LRL \rightarrow LAB \rightarrow  LLN) $\\
6 & 1.32200 - 1.32280 & $ LLNLAB~~(LLNLLN \rightarrow  LLNLAB \rightarrow
LLNLRL) $\\
\hline
\end{tabular}
\end{center}
\end{table}                      

\section{Propulsion aspects}
\label{prop}
\begin{figure}
\centerline{\epsfig{file=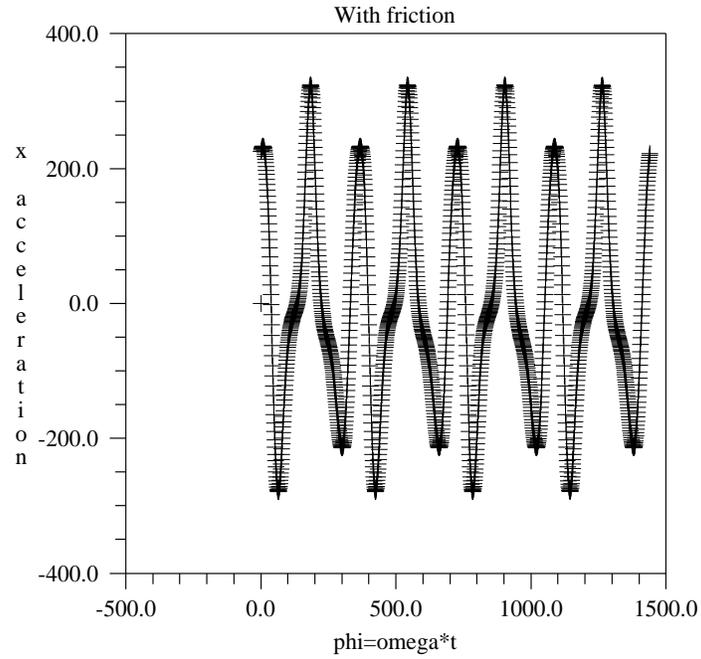,height=3.7in,width=3.7in}}
\caption{Acceleration in the $x$ direction of the center--of--mass
of the rod as a function of time (low friction).}\label{xacc1}
\end{figure}

\begin{figure}
\centerline{\epsfig{file=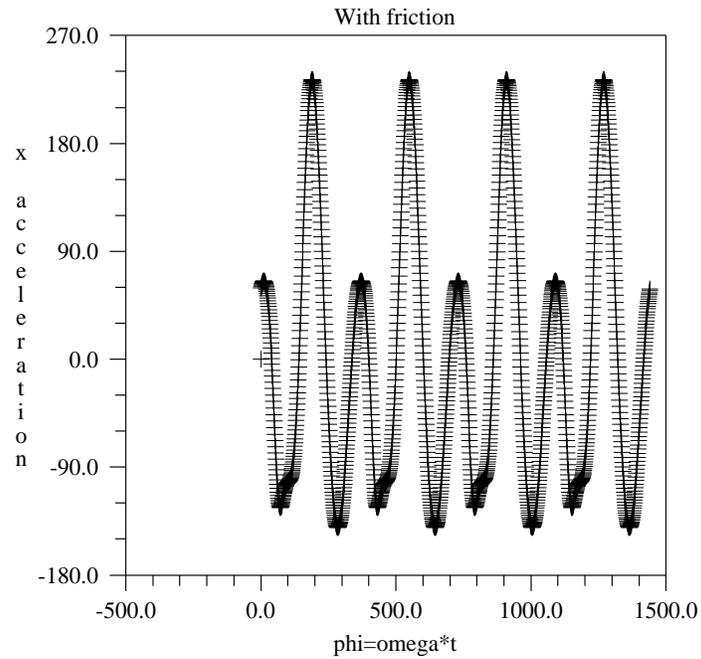,height=3.7in,width=3.7in}}
\caption{Acceleration in the $x$ direction of the center--of--mass
of the rod as a function of time (high friction).}\label{xacc2}
\end{figure}

Apart from its fairly rich non--linear structure, another
interesting aspect of the model considered here is its
use in a mechanical propulsion device.
To see these aspects, let us use Eqs.~(\ref{eq:dxdy}, \ref{eq:coord})
to solve for the coordinates (in Cartesian space) of the
center of mass of the rod. We are particularly interested 
in this section in the acceleration; two representative
graphs of the acceleration in the $x$ direction are shown
in Figs.~\ref{xacc1} and \ref{xacc2}.
Similar results hold for the acceleration in the $y$--direction.
These graphs were generated with a particular set of parameters
and initial conditions 
of the model whose precise values are not important for our
purposes here. The interesting aspect is that the acceleration
in, say, the $x$ direction shows a definite bias in 
favor of being larger in magnitude in for, in this case,
positive values over negative values.
\par
Let us imagine that we place this mechanical model on a cart
which is resting on a surface
and let it begin its motion. The acceleration of the rod
will, by Newton's 3$^{\rm rd}$ law, cause a back reaction,
causing the cart to oscillate back and forth. First assume
the surface is frictionless. Even though the acceleration of the
rod is larger in one particular direction than the opposite
one, the cart will not show any net movement -- the larger
acceleration in, say, the $+x$ direction is compensated by
the smaller acceleration in the $-x$ directions which
lasts for comparatively longer periods of time. However, let
us now imagine that there is some friction between the cart
and the surface. It is possible, with the right combination
of parameters and initial conditions, 
that this friction ``absorbs'' the back reaction
in the direction for which the acceleration is smallest, but
doesn't completely absorb the back reaction in the other
direction. In this case the cart can ``push off'' the surface,
and exhibit a net motion.
\par
Similar effects will hold for motion in the $y$ direction,
and so the net motion will be a combination of motion
in both directions. However, if we place two such mechanical
devices on the cart and have them rotate in opposite directions,
the back reactions in, say, the $y$ direction can be arranged to
cancel, leaving a net motion in only the $x$ direction.
Such a device has been constructed
(U.~S.~Patent \#4,631,971 and U.~S.~Patent Application \#268,914);
one rather striking illustration of its operation
is that the device, self--contained within a box, when
placed in a canoe in a swimming pool, will propel the canoe
in one particular direction, without the need for a propeller
in contact with the water.
\par
Although perhaps surprising, the principle behind this device
is intuitively known to many children. Have
a child sit on a piece of cardboard 
on a smooth floor and then tell her to move forward by rocking. 
If one watches closely, the child will rock in one direction quickly 
and return in the other direction slowly. The change in momentum of 
the child leads to a reactive force being transferred to the cardboard, 
but the uneven rocking results in this force overcoming the frictional 
force in only one particular direction. The net result
is that the child will propel herself in one direction. 
\section{Conclusions}
\label{con}
In this paper we have studied a simple mechanical model
of a periodically driven nonlinear mechanical system.
Bifurcation diagrams, dimensions, power spectra, and
Lyapunov exponents were derived in order to
locate regions of quasiperiodic, 
periodic and chaotic behavior within the parameter space of the
system. Within the parameter space is the
the coexistence of a chaotic attractor and a limit cycle, which implies that 
initial conditions may determine whether the system will exhibit periodic or 
chaotic behavior. We also discussed some aspects of the
model as it relates to a propulsion device.
This system, although relatively simple,
exhibits a rich behavior with a number of interesting
non--linear effects being present.
\par
We have also performed a symbolic analysis of the model.
By constructing the proper Poincar\'e section in the phase 
space for a system of ODEs,
the symbolic dynamics can be constructed based on the appropriate
partitioning
of the phase portrait and it turns out to be an 
efficient and powerful way to
explore the global properties of the system both in the phase and parameter
spaces. 
Up to now, the symbolic dynamics has been applied to the analysis of the 
NMR--laser chaos model \cite{zl95,lwz96}, the two--well Duffing
equation \cite{xzh95}, the forced Brusselator \cite{lz95,lzh96}, and 
the Lorenz model \cite{fh96,zl97,hlz98},
 in addition to the system discussed in this paper. 
Along a certain direction in the parameter space this model exhibits
various properties, such as, periodicity, quasiperiodicity, chaos, 1D and 2D
features,
etc. In some other directions or regions of the parameter space the
model would
also display more or less similar behavior. We have established the
3-letter symbolic 
dynamics for the model and found that the ordering rules of 
sequences, 
the forced Brusselator in the regime of annular dynamics and the dissipative
standard 
map at some parameters are the same. As a matter of fact, the NMR-laser
chaos model, 
the forced Brusselator in the regime of interval dynamics and the H\'enon
map with 
a positive Jacobian also have similar 2-letter symbolic dynamics and share
the same 
ordering rules of sequences. It, therefore is meaningful in a sense to
classify 
the systems of ODEs according to their ordering rules of sequences. The
ODEs 
investigated under the guidance of symbolic dynamics to day are quite
limited though.  

\section*{Acknowledgements}
This work was supported by the Natural Sciences and Engineering
Research Council of Canada.

\end{document}